\documentclass[aip,pop,reprint]{revtex4-2}
\usepackage[bookmarksnumbered=true,hidelinks]{hyperref}
\usepackage[utf8]{inputenc}
\usepackage[english]{babel}
\usepackage{bm}
\usepackage{amsmath}
\usepackage[shortcuts]{extdash}
\usepackage{graphicx}
\usepackage{tabularx}
\usepackage{soul}
\usepackage[dvipsnames,table,xcdraw]{xcolor}

\definecolor{COMMENT_GREEN}{HTML}{4CBB17}
\definecolor{COMMENT_GREEN_DIM}{HTML}{228B22}

\usepackage[most]{tcolorbox}
\newtcolorbox{Comment}{colback=COMMENT_GREEN!10,colframe=COMMENT_GREEN_DIM!30,fonttitle=\bfseries,title=\textcolor{white}{\large Comment $\vphantom{\int}$}, boxrule=0mm, arc=0mm,colbacktitle=COMMENT_GREEN_DIM, leftrule = 3mm}

\hypersetup{
    colorlinks = true,
    linkcolor = blue,
    anchorcolor = blue,
    citecolor = blue,
    filecolor = blue,
    urlcolor = blue
}

\DeclareUnicodeCharacter{2010}{-}

\renewcommand{\boldsymbol}[1]{\bm{#1}}

\def\Energy{\mathcal{E}}
\def\SpectralEnergy{\mathcal{I}}
\def\SheathPotential{\Phi}
\def\FluxProbabilityFunction{F}

\def\ICTSR{\SpectralEnergy_{\rm CTSR}}

\def\eps{\varepsilon}
\def\tret{t_{\rm ret}}

\def\deg{\ensuremath{^\circ}}

\def\gammaAvg{\langle \gamma_h \rangle}

\def\nameFront{\mathrm{f}}
\def\nameRear{\mathrm{e}}

\def\nameTraj{\mathrm{r}}
\def\nameRealParticle{-}
\def\nameImageParticle{+}
\def\nameMultipleParticles{\pm}

\def\r{\boldsymbol{r}}

\def\p{\boldsymbol{p}}

\def\u{\boldsymbol{u}}

\def\j{\boldsymbol{j}}

\def\fatbeta{\boldsymbol{\beta}}
\def\j{\boldsymbol{j}}
\def\E{\boldsymbol{E}}

\def\A{\boldsymbol{A}}

\def\AInterfaces{\bar{\A}_{\rm TR}\left(\rhat, \nu\right)}

\def\nubar{\bar{\nu}}

\def\tbar{\bar{t}}
\def\taubar{\bar{\tau}}
\def\Deltatbar{\overline{\Delta t}}
\def\xibar{\bar{\xi}}
\def\fatxibar{\bar{\boldsymbol{\xi}}}
\def\DeltaRetBar{\Deltatbar_{\nameTraj}}
\def\DeltaRet{\Delta t_{\nameTraj}}
\def\psiphiIntegral{\sideset{}{_{\psi,\varphi}} \iint}

\def\Rhat{\boldsymbol{\hat{R}}}
\def\rhat{\boldsymbol{\hat{r}}}
\def\zhat{\boldsymbol{\hat{z}}}

\def\dd{\mathrm{d}}

\def\dpa{\partial}

\newcommand{\retardedTime}[2][\tret]{\left[ #2 \right]_{#1}}

\newcommand{\seccite}[1]{Sec.~#1}
\newcommand{\figcite}[1]{Fig.~#1}
\newcommand{\appcite}[1]{Appendix~#1}

\DeclareFontFamily{U}{mathx}{\hyphenchar\font45}
\DeclareFontShape{U}{mathx}{m}{n}{<-> mathx10}{}
\DeclareSymbolFont{mathx}{U}{mathx}{m}{n}
\DeclareMathAccent{\widebar}{0}{mathx}{"73}

\newcommand{\textReferee}{\textcolor{Black}}

\begin{document}
\DeclareGraphicsExtensions{.pdf,.png,.jpg}

\title{Modeling terahertz emissions from energetic electrons and ions in foil targets irradiated by ultraintense femtosecond laser pulses}
\author{E. Denoual}
\email{emilien.denoual@cea.fr}
\affiliation{CEA, DAM, DIF, F-91297 Arpajon, France}
\affiliation{Universit\'e Paris-Saclay, CEA, LMCE, F-91680 Bruy\`eres-le-Ch\^atel, France}
\author{L. Berg\'e}
\email{luc.berge@cea.fr}
\affiliation{CEA, DAM, DIF, F-91297 Arpajon, France}
\affiliation{Universit\'e Paris-Saclay, CEA, LMCE, F-91680 Bruy\`eres-le-Ch\^atel, France}
\affiliation{Centre des Lasers Intenses et Applications, Université de Bordeaux-CNRS-CEA, F-33405 Talence Cedex, France}
\author{X. Davoine}
\email{xavier.davoine@cea.fr}
\affiliation{CEA, DAM, DIF, F-91297 Arpajon, France}
\affiliation{Universit\'e Paris-Saclay, CEA, LMCE, F-91680 Bruy\`eres-le-Ch\^atel, France}
\author{L. Gremillet}
\email{laurent.gremillet@cea.fr}
\affiliation{CEA, DAM, DIF, F-91297 Arpajon, France}
\affiliation{Universit\'e Paris-Saclay, CEA, LMCE, F-91680 Bruy\`eres-le-Ch\^atel, France}

\date{\today}
\begin{abstract}
Terahertz (THz) emissions from fast electron and ion currents driven in relativistic, femtosecond laser-foil interactions are examined theoretically. We first consider the radiation from the energetic electrons exiting the backside of the target. Our kinetic model takes account of the coherent transition radiation due to these electrons crossing the plasma-vacuum interface as well as of the synchrotron radiation due to their deflection and deceleration in the sheath field they set up in vacuum. After showing that both mechanisms tend to largely compensate each other when all the electrons are pulled back into the target, we investigate the scaling of the net radiation with the sheath field strength. We then demonstrate the sensitivity of this radiation to a percent-level fraction of escaping electrons. We also study the influence of the target thickness and laser focusing. The same sheath field that confines most of the fast electrons around the target rapidly sets into motion the surface ions. We describe the THz emission from these accelerated ions and their accompanying hot electrons by means of a plasma expansion model that allows for finite foil size and multidimensional effects. Again, we explore the dependencies of this radiation mechanism on the laser-target parameters. Under conditions typical of current ultrashort laser-solid experiments, we find that the THz radiation from the expanding plasma is much less energetic -- by one to three orders of magnitude -- than that due to the early-time motion of the fast electrons.
\end{abstract}
\maketitle

%


\section{Introduction}
\label{sec:introduction}

Intense sources of terahertz (THz) radiation are drawing growing interest as their oscillation period, of the order of a few picoseconds, makes them ideally suited for the study of numerous phenomena evolving on similar time scales~\cite{Tonouchi:np:2007}. Their main direct applications include medical imaging~\cite{Pickwell:jpd:2006}, molecular spectroscopy~\cite{, Mittleman:apb:1999, Berge:epl:2019}, tomography~\cite{Chan:rop:2007}, and modification of condensed matter properties~\cite{Orenstein:sci:2000, Kampfrath:np:2013, Li:sci:2019}, to cite only a few. While intense lasers offer promising prospects for developing compact, ultrashort THz sources, the main challenge nowadays is to produce broadband THz pulses with $\rm mJ$-level energies as is required for various uses \cite{Andreasen:pre:2012, Nanni:nc:2015, Li:sci:2019, Novelli:mat:2020, Vella:sa:2021}.
This is a nontrivial task as the most widely explored THz generation mechanisms, namely, optical rectification in asymmetric crystals~\cite{hebling:oe:2002, Vicario:prl:2014, yeh:apl:2007} or photoionization of gases by two-color, moderate-intensity ($\lesssim 10^{15}\,\rm W\,cm^{-2}$), femtosecond laser pulses~\cite{Kim:np:2008, Clerici:prl:2013, Nguyen:pra:2018}, are to date limited to tens of $\rm \mu J$ THz pulse energies and $\sim 1\,\rm GV\,m^{-1}$ field strengths.

A more auspicious approach is to irradiate gaseous targets at relativistic laser intensities ($I_L > 10^{18}\,\rm W\,cm^{-2}$). In this regime, it has been demonstrated that coherent transition radiation (CTR) from wakefield-accelerated relativistic electron bunches at the rear plasma boundary can lead to intense THz emissions, characterized by a few $100\,\rm \mu J$ energy yield and $> 10\,\rm GV m^{-1}$ field strength~\cite{Leemans:prl:2003,Dechard:prl:2018}. Such a radiation is coherent because the typical dimensions of the electron bunches ($\sim 1-5 \,\rm \mu m$) are smaller than the THz radiation wavelengths ($> 10-100\,\rm \mu m$). Consequently, the THz pulse energy essentially scales as the square number of fast electrons, which makes it a potentially very efficient mechanism~\cite{Schroeder:pre:2004}. 

CTR also operates in relativistic laser-solid interactions, whereby, compared to gas targets, it benefits from a stronger absorption of the laser energy into MeV-range electrons, and hence from an increased number of radiating particles \cite{Ding:apl:2013, Ding:pre:2016, Liao:prl:2016, Jin:pre:2016, Herzer:njp:2018, Woldegeorgis:pre:2019, Liao:pnas:2019, Liao:prx:2020, Dechard:pop:2020}. However, because different acceleration mechanisms are at play \cite{Kemp:nf:2014}, these energetic electrons are generally characterized by a much larger ($\sim 100\times$) angular divergence than those generated by laser wakefields, which translates into a broader CTR emission cone. Yet, owing to its high density ($\sim 10^{19-21}\,\rm cm^{-3}$), the hot-electron population does not only radiate via CTR when exiting a solid foil.

\begin{figure}
    \centering
    \includegraphics[width=0.865\linewidth]{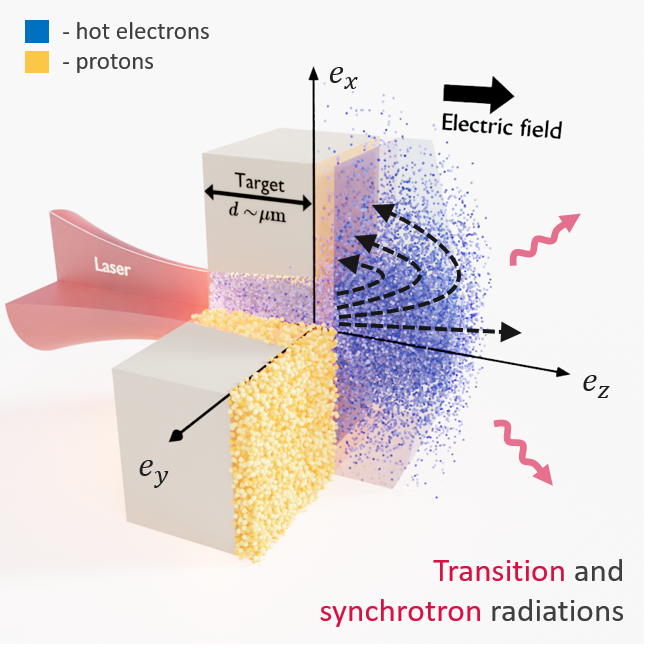}
    \includegraphics[width=0.865\linewidth]{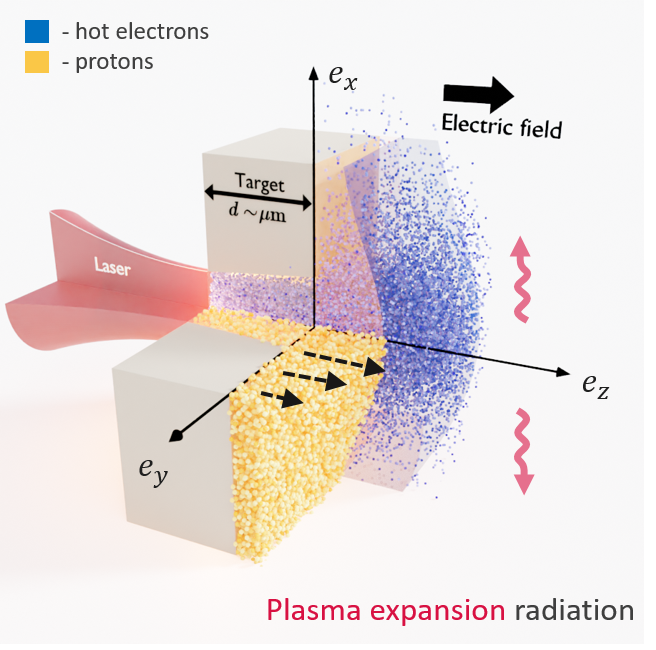}
    \caption{The two major THz radiation mechanisms in relativistic laser-foil interactions considered in this work. (top) First, coherent transition and synchrotron radiations are generated by the laser-accelerated electrons crossing the target backside and being reflected (or not) in the sheath field they set up in vacuum. (bottom) Subsequently, the nonneutral layers at the edges of the rear-side expanding plasma emit a dipole-type radiation.}
    \vspace{-1.0em}
    \label{fig:ctr:illustration:CTSR_PER}
\end{figure}

The latter mechanism indeed assumes that the fast electrons propagate ballistically across the plasma-vacuum interface whilst most of them actually get reflected in the strong charge-separation field that they set up in vacuum \cite{Fill:pop:2001, Link:pop:2011, Rusby:hplse_2019}. This results in an additional coherent, synchrotron-type radiation (CSR) of polarity opposite to that of CTR \cite{Liao:pnas:2019, Liao:prx:2020}. An additional complication follows from the fraction of fast electrons that are able to escape the target, and thus just emit a single burst of CTR. The net THz radiation resulting from those combined processes, CTR and CSR, will be referred to as CTSR [Fig.~\ref{fig:ctr:illustration:CTSR_PER}(top)].

The sheath electric field induced by the hot electrons on both sides of the target subsequently sets into motion the surface ions, a process widely known as target normal sheath acceleration (TNSA) in the context of relativistic laser-plasma interactions~\cite{Wilks:pop:2001, Mora:prl:2003, Macchi:rmp:2013}. Because of their highest charge-to-mass ratio, the protons, generally present in the form of contaminants, react the fastest to that field, reaching velocities $\sim 0.1c$ ($c$ is the speed of light) on $\sim 1\,\rm ps$ timescales. The resultant expanding plasma at the target backside, composed of accelerated ions and electrostatically trapped hot electrons, comprises two charge-separation regions: one negatively charged at its outer boundary, and one positively charged around its inner boundary \cite{Mora:prl:2003}. Their time-varying properties lead to a dipole-type, low-frequency radiation
\cite{Gopal:prl:2013, Gopal:ol:2013, Herzer:njp:2018, Woldegeorgis:pre:2019}, henceforth labeled plasma expansion radiation (PER) [Fig.~\ref{fig:ctr:illustration:CTSR_PER}(bottom)].

Here, we develop two models to estimate the THz radiated spectra and energy yields from the two aforementioned mechanisms, CTSR and PER, based, respectively, on the fast-electron dynamics alone at the unperturbed target backside and, on slower time scales, fast-electron-induced ion acceleration into vacuum. Compared to previous related works \cite{Schroeder:pre:2004, Woldegeorgis:pre:2019, Liao:pnas:2019, Liao:prx:2020}, we propose a unified kinetic treatment of the destructively interfering CTR and CSR as resulting from the beam electrons' trajectories in vacuum. Moreover, we model PER using a refined description of ion acceleration, notably allowing for the time-decreasing surface charge density in the expanding sheath and for the hot-electron cooling in thin foil targets. We expect our modeling to be mostly valid in the case of micrometer-range foil targets driven by relativistic femtosecond laser pulses. Our main prediction is that, under such conditions, the energy radiated by the sole fast electrons via CTR and CSR should exceed that due to plasma expansion by at least tenfold, and more in the likely case of a percent-level fraction of escaping electrons. Our work thus settles a lingering debate on the dominant THz radiation process in ultrashort-pulse laser experiments  \cite{Gopal:prl:2013, Gopal:ol:2013, Jin:pre:2016, Liao:prl:2016, Herzer:njp:2018, Woldegeorgis:pre:2019, Liao:prx:2020}.

We start this paper by outlining the framework of our study in Sec.~\ref{subsec:ctr:framework}. In Sec.~\ref{subsec:ctr:one_electron_radiation}, we recall the basics of far-field radiation from a charged particle moving in vacuum near a perfect conductor, using the method of the image charge. Section~\ref{subsec:ctr:one_electron_trajectory} then characterizes the radiation from a single electron exiting a perfect conductor and experiencing a constant electric field in vacuum. In Sec.~\ref{subsec:ctr:electron_beam}, this problem is generalized to the case of an energy-angle distributed electron beam originating from the laser-irradiated side of the target. The respective integral expressions of CTR and CSR are then detailed. In Sec.~\ref{subsec:ctr:parameters}, we specify the various parameters of the model, of relevance to femtosecond laser-foil interactions. 

Section~\ref{subsec:ctr:interplay} presents the main spectral features of CTR, CSR and CTSR in a typical ultrashort laser-foil configuration. A major finding is that CTR and CSR closely compensate each other in the THz domain, when all fast electrons are made to reflect back into the target. The scaling of the CTSR yield with the sheath field strength is examined in Sec.~\ref{subsec:ctr:return_time}, while the possibly dominant contribution of even a small ($\sim 1\,\%$) fraction of escaping electrons is discussed in Sec.~\ref{subsec:ctr:escaping_electrons}. The sensitivity of the net THz radiation to target thickness and laser focusing is addressed in Secs.~\ref{subsec:ctr:target thickness} and \ref{subsec:ctr:laser_field}.

Section~\ref{sec:sr:radiation_expanding_plasma} next considers the radiation arising from the plasma expansion. The general formalism of our approach is presented in Sec.~\ref{subsec:sr:formalism}. After detailing the underlying model of TNSA in Sec.~\ref{subsec:sr:tnsa_regimes}, we derive the integral form of the PER energy spectra in Sec.~\ref{subsec:sr:energy_spectrum}. The dependencies of PER on the system's parameters are investigated in Secs.~\ref{subsec:sr:laser_parameters} and \ref{subsec:sr:target_thickness}. Finally, Sec.~\ref{sec:conclusions} gathers our conclusions. 

%

\section{Coherent transition and synchrotron radiations from fast electrons}
\label{sec:ctr:transition_and_synchrotron_radiations}

\subsection{Framework of the study}
\label{subsec:ctr:framework}

The system investigated consists of a thin ($d \sim 1\,\rm \mu m$) solid foil impacted by an ultraintense ($I_L \sim 10^{20}\,\rm W\,cm^{-2}$) and ultrashort ($\tau_L \sim 30~\rm fs$) laser pulse of wavelength $\lambda_L \sim 1~\rm \mu m$, focused to a few-$\mu \rm m$ spot size ($w_L$). We suppose that the target is fully ionized within a few optical cycles, hence turning into a plasma of overcritical electron density $n_{e0} \gg n_c$ (where $n_c = 4 \pi^2 \epsilon_0 m_e c^2/ e^2 \lambda_L^2$ is the critical density, $m_e$ the electron mass, $e$ the elementary charge and $ \epsilon_0$ the vacuum permittivity), treated hereafter as a perfect conductor with sharp boundaries.

The relativistic electrons driven by the laser pulse at the target front side form a bunch of typical length $c\tau_L \sim 10\,\rm \mu m$, width $w_L$, number density $n_h \sim n_c$ and average Lorentz factor $\langle \gamma_h \rangle$. They are assumed to propagate ballistically on their first pass through the material. Upon crossing the backside of the target, they undergo an abrupt change in permittivity which causes transition radiation~\cite{Mikaelian:book:1972, Ginzburg:book:1979, *Ginzburg:ps:1982}. Wavelengths larger than the bunch size, i.e., lying in the THz domain, are emitted coherently~\cite{Zheng:pop:2003, Schroeder:pre:2004, Bellei:ppcf:2012}.

When exiting the target, the hot electrons induce a strong electric sheath field, parallel to the surface normal and of typical strength $E_0 \sim (m_e c \omega_0/e)\sqrt{\langle \gamma_h \rangle (n_h/n_c)} \sim 10^{12-13}\,\rm V\,m^{-1}$~\cite{Fill:pop:2001, Mora:prl:2003, Ridgers:pre:2011}, which pulls the vast majority of the accelerated electrons back into the foil~\cite{Rusby:hplse_2019}. The synchrotron emission that they generate while experiencing the sheath field is another source of coherent THz radiation. 
One can anticipate, as will be examined in detail below, that this deceleration-induced radiation will interfere destructively with CTR, hereafter interpreted as resulting from the apparent sudden acceleration of the electrons at the conductor's surface~\cite{Ginzburg:book:1979, *Ginzburg:ps:1982}.

The problem of CTR from laser-generated electron beams crossing at constant velocity a plasma-vacuum boundary has already been widely addressed theoretically in multiple frequency domains~\cite{Zheng:pop:2003, Schroeder:pre:2004, Bellei:ppcf:2012}. However, with the exception of Refs.~\onlinecite{Liao:pnas:2019, Liao:prx:2020}, the simultaneous modeling of CTR and CSR due to the non-ballistic motion of beam electrons in vacuum has received little attention so far. As we will see further on, when almost all of these electrons are made to reflux into the target, CTR and CSR produce THz fields with very similar spectra, yet of opposite polarity, which therefore tend to cancel each other out. The net THz radiation is thus determined by an integral over the electron trajectories in vacuum, which depend on the initial electron energy and propagation angle.

This radiation, as will be shown in Sec.~\ref{subsec:ctr:escaping_electrons}, is also highly sensitive to the non-compensated CTR from the fraction of high-energy electrons able to escape the target permanently. These, believed to make up at most a few percent of the whole hot-electron population~\cite{Rusby:hplse_2019}, can escape the target at near the speed of light and thus contribute to a single uncompensated CTR flash~\cite{Liao:pnas:2019, Liao:prx:2020}. The interplay of those mechanisms, and their variations with the laser-target parameters, will be thoroughly examined in Secs.~\ref{subsec:ctr:target thickness} and \ref{subsec:ctr:laser_field}.

Before proceeding, two limitations of our modeling are already worth mentioning. First, we will consider only the first excursion of the fast electrons into vacuum, and thus neglect their subsequent THz emissions while they bounce back and forth across the target ~\cite{Ding:pre:2016, Jin:pre:2016, Dechard:pop:2020}.
Accordingly, the following estimates of the energy radiated from the sole fast electrons should be considered as lower values.  
Second, by assuming a target of infinite transverse size, we will discard the multiple THz emissions that are expected to arise from the laser axis and the target edges when the fast electrons recirculate transversely across a finite-width target~\cite{Zhuo:pre:2017, Dechard:pop:2020}.  

\subsection{Radiation from an electron accelerated in vacuum near a perfect conductor}
\label{subsec:ctr:one_electron_radiation}

The electromagnetic field radiated in vacuum by an accelerated charged particle can be split into two components~\cite{Jackson1999, Zangwill}: a \textit{velocity field} that rapidly decays in space as $R^{-2}$ and an \textit{acceleration field} that decays as $R^{-1}$, where $R = \left\vert \r - \r_p(t)\right\vert$ is the distance from the particle, located at $\r_p(t)$, to the detector, located at $\r$. In the far-field limit, the detected field reduces to the acceleration field \cite{Jackson1999, Zangwill},
\begin{equation}
    \E_{\rm acc}(\r,t) = \frac{q}{4 \pi c \eps _0} \retardedTime{\frac{\Rhat \times \big\{ (\Rhat - \fatbeta_p) \times \dot{\fatbeta}_p \big\} }{R \big(1-\fatbeta_p \cdot \Rhat \big)^3}}\,,
    \label{equ:ctr:one_electron:acceleration_field}
\end{equation}
where $\fatbeta_p(t) = \dot{\r}_p(t)/c$ is the normalized velocity of the particle, $\dot{\fatbeta}_p(t)$ its normalized acceleration, $q$ its charge, $\eps_0$ the permittivity of free space and $\Rhat = \left( \r - \r_p(t)\right)/R$ the unit direction of observation. The operator $\retardedTime{\,.\,}$ evaluates its argument at the retarded time $t_{\rm ret}$, implicitly defined by $t_{\rm ret} = t - R(\tret)/c$. 

\begin{figure}
    \centering
    \includegraphics[width=0.9\linewidth]{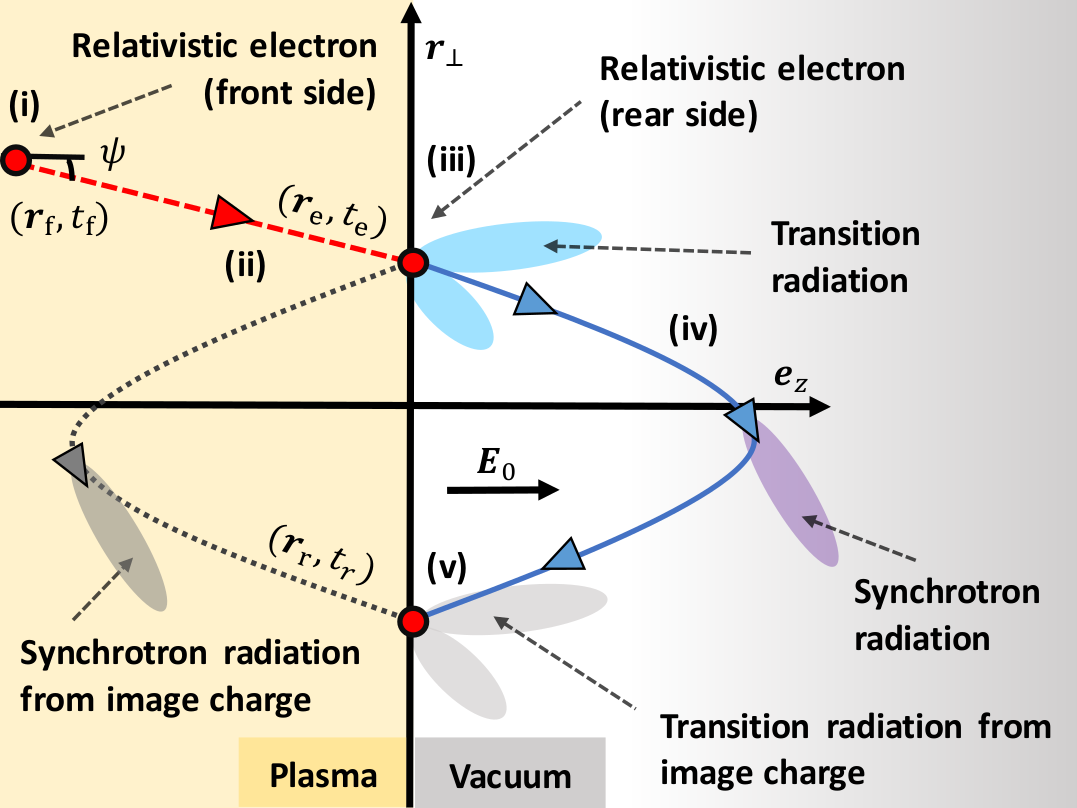}
    \caption{Transition and synchrotron radiations from a relativistic electron confined at the target-vacuum interface. The electron trajectory starts at the left (laser-irradiated) side of the target, at time $t_{\nameFront}$ and position $\r_{\nameFront}$. The initial electron momentum is $\p_{\nameRear} = p_{\nameRear} (\sin \psi \cos \varphi, \sin \psi \sin \varphi, \cos \psi)$ where $\psi$ and $\varphi$ are the polar and azimuthal angles, respectively. After traveling ballistically through the target (red dashed curve), the electron crosses its rear side at time $t_{\nameRear}$ and position $\r_{\nameRear}$. In vacuum, it is then reflected by the longitudinal sheath field $\boldsymbol{E}_0$, assumed homogeneous and stationary, and re-enters the target at time $t_{\nameTraj}$ and position $\r_{\nameTraj}$ (blue solid curve). The color gradient represents the electrostatic potential profile $\SheathPotential(z)$, see Sec.~\ref{subsec:ctr:one_electron_trajectory}.
    The trajectory of the associated image charge is plotted as a grey dotted curve.
    The observation direction is $\rhat = (\sin \theta \cos \Psi, \sin \theta \sin \Psi, \cos \theta) = \r/r$, where $\theta$ and $\Psi$ are the corresponding polar and azimuthal angles. In this 2D representation, for simplicity, the observer is taken to lie in the plane of particle motion (i.e. $\Psi = \varphi$).}
    \vspace{-1.0em}
    \label{fig:ctr:illustration:target_thin_foil}
\end{figure}

When the particle moves in vacuum in the vicinity of a perfect conductor, Eq.~\eqref{equ:ctr:one_electron:acceleration_field} must be corrected so that the tangential component of the electric field vanishes at the conducting surface (assumed planar). Microscopically, this arises because of the additional field generated by polarization surface currents \cite{Shkvarunets:prab:2008}. \textReferee{The same effect is obtained by replacing the conductor by an ``image'' particle of charge $-q$, moving symmetrically to the real particle on the other side of the boundary, and stopping instantaneously when they meet (see Fig.~\ref{fig:ctr:illustration:target_thin_foil}). Since the two configurations yield identical fields in the semi-infinite space occupied by the real particle, the radiation produced in the vacuum by the polarization currents is identical to the radiation from the fictitious image particle} \cite{Ginzburg:book:1979, *Ginzburg:ps:1982, Carron:jewa:2000, Bolotovskii:UP:2009}. 
Hence, the total radiation from the particle in the presence of the conductor is
\begin{equation}
    \label{equ:ctr:sum_of_outgoing_charge_and_image_charge}
    \E_{\rm rad}(\r,t) = \E_{\rm acc}^{\nameRealParticle}(\r,t) + \E_{\rm acc}^{\nameImageParticle}(\r,t)\,,
\end{equation}
where $\E_{\rm acc}^{\nameRealParticle}(\r,t)$ and $\E_{\rm acc}^{\nameImageParticle}(\r,t)$ represent the acceleration fields generated by, respectively, the real (superscript $^\nameRealParticle$) and image (superscript $^\nameImageParticle$) particles. Both fields are evaluated at their own retarded times, $t_{\rm ret}^{\nameRealParticle}$ or $t_{\rm ret}^{\nameImageParticle}$, depending on the positions of the particle and observer.

The power radiated by the particle and its image charge per unit solid angle reads~\cite{Zangwill, Jackson1999}
\begin{equation}
    \label{equ:ctr:power_distribution}
    \frac{d P_{\rm rad}(\rhat, t)}{d \Omega} = c \eps_0 \left\vert \retardedTime[t_{\rm ret}^{\nameRealParticle}]{R^\nameRealParticle \E_{\rm acc}^\nameRealParticle} + \retardedTime[t_{\rm ret}^{\nameImageParticle}]{R^\nameImageParticle \E_{\rm acc}^\nameImageParticle} \right\vert^2 \,,
\end{equation}
where we introduce the direction of observation $\rhat = (\sin \theta \cos \Psi, \sin \theta \sin \Psi, \cos \theta) = \r/r$, the polar angle $\theta$, the azimuthal angle $\Psi$, and $\dd \Omega = \sin \theta \dd \theta \dd \Psi$. The total energy radiated per unit solid angle can be expressed as~\cite{Jackson1999, Zangwill}:
\begin{equation}
    \label{equ:ctr:definition:energy_distribution}
    \frac{\dpa \Energy_{\rm rad}(\rhat)}{ \dpa \Omega} = \int_{-\infty}^{\infty} \frac{\dpa P_{\rm rad}(\rhat, t)}{\dpa \Omega} \dd t \equiv \int_0^\infty  \frac{\dpa^2 \SpectralEnergy_{\rm rad}(\rhat, \nu)}{\dpa \nu \dpa \Omega} \dd \nu \,,
\end{equation}
where the second equality defines $\partial^2 \SpectralEnergy_{\rm rad}/\partial \nu \partial \Omega$, the energy radiated per unit solid angle and frequency ($\nu$) interval. Upon combining Eqs.~\eqref{equ:ctr:one_electron:acceleration_field}-\eqref{equ:ctr:definition:energy_distribution}, performing the change of variable $t \mapsto \tret$ and assuming that the accelerated particle remains far away from the observer, i.e., $R(t) \simeq r - \rhat \cdot \r_p(t)$, one obtains~\cite{Jackson1999, Zangwill}
\begin{widetext}
\begin{equation}
    \label{eq:frequency-angle_spectrum_single_electron}
    \frac{\partial^2 \SpectralEnergy_{\rm rad}}{\partial \nu \partial \Omega}(\rhat, \nu) = \frac{q^2}{8 \pi^2 \eps_0 c} \left\vert \int 
    \frac{\rhat \times [ (\rhat -\fatbeta_p^-) \times \dot{\fatbeta}_p^{-} ] }{(1 - \fatbeta_p^- \cdot \rhat )^2}
    e^{-2i\pi \nu (t - \rhat \cdot \r_p^-(t)/c )} - \frac{\rhat \times [ (\rhat - \fatbeta_p^+) \times \dot{\fatbeta}_p^{+} ] }{(1 - \fatbeta_p^+ \cdot \rhat )^2}
    e^{-2i\pi \nu (t - \rhat \cdot \r_p^+(t)/c )} \,\dd t
    \right\vert^2\,.
\end{equation}
\end{widetext}
This expression describes the full energy spectrum radiated by a charged particle moving in vacuum in the vicinity of a perfect conductor. \textReferee{In the following, we will consider the case of electron trajectories that start at the conductor surface and return to it after some time spent in the vacuum.}

\subsection{Transition and synchrotron radiations from an electron exiting and coming back into a perfect conductor}
\label{subsec:ctr:one_electron_trajectory}

The generic scenario addressed in our study is sketched in Fig.~\ref{fig:ctr:illustration:target_thin_foil}. It consists of an electron (i) laser-accelerated at the front side of a foil target, (ii) traveling ballistically through it, (iii) crossing its rear side, (iv) being reflected in vacuum by the sheath field established by the hot electrons and (v) re-crossing the target's rear surface. Because of plasma shielding, stages (i) and (ii) of the electron motion do not lead to outgoing radiation from the backside of the target. By contrast, stages (iii) and (v), corresponding to apparent sudden accelerations of the electron and its image charge, give rise to two consecutive flashes of transition radiation \cite{Mikaelian:book:1972, Ginzburg:book:1979, *Ginzburg:ps:1982}, at the exit ($t=t_{\nameRear}$) and return ($t=t_{\nameTraj}$) times, while stage (iv) generates synchrotron-type radiation over the time interval $t_{\nameRear} < t < t_{\nameTraj}$. \textReferee{Although it will be shown below that this distinction between transition and synchrotron radiations is often rather artificial due to their interfering behavior, it has pedagogical value in helping make connection with previous works.}

To identify the transition- and synchrotron-type components in Eq.~\eqref{eq:frequency-angle_spectrum_single_electron},
where the time integral is performed over the particle trajectory in vacuum, it is convenient to express the normalized velocities ($\fatbeta_p^\pm$) of the electron and its image charge as
\begin{equation}
\label{equ:ctr:one_electron:acceleration_profile}
    \fatbeta_p^\pm(t) =
    \begin{cases}
      \fatbeta_{\nameRear}^\pm \left[ \frac{t-t_{\nameRear}}{\delta t}+1 \right] &\,  t_{\nameRear}-\delta t < t \le t_{\nameRear}  \,, \\
      \fatbeta^\pm(t) &\,   t_{\nameRear} < t < t_{\nameTraj} \,,\\
      \fatbeta_{\nameTraj}^\pm \left[ \frac{t_{\nameTraj} - t}{\delta t} + 1 \right] &\,  t_{\nameTraj} \le t < t_{\nameTraj} + \delta t \,,\\
      0 &\, \text{otherwise}\,,
    \end{cases}\medskip
\end{equation}
where $\fatbeta_{\nameRear}^{\pm} \equiv \fatbeta^{\pm} (t_{\nameRear})$ and  $\fatbeta_{\nameTraj}^{\pm} \equiv \fatbeta^{\pm} (t_{\nameTraj})$ denote the velocities at the exit and return times, respectively, and $\delta t$ is an infinitesimal time interval.
\textReferee{The latter is introduced to describe the sudden appearance -- or disappearance -- of the particle across the boundary of the perfectly shielding target, and will enable us to easily generalize the standard formula for the transition radiation spectrum \cite{Ginzburg:book:1979, *Ginzburg:ps:1982} in the case of two consecutive surface crossings.}

Noting that \cite{Jackson1999}
\begin{equation}
   \frac{\rhat \times [ (\rhat - \fatbeta_p^\pm) \times \dot{\fatbeta}_p^\pm ] }{\left( 1 - \fatbeta_p^\pm \cdot \rhat \right)^2 } = \frac{\dd}{\dd t}\left[ \frac{\rhat \times (\rhat \times \fatbeta_p^\pm)}{1-\fatbeta_p^\pm \cdot \rhat} \right] \,,
\end{equation}
we can perform an integration by parts in Eq.~\eqref{eq:frequency-angle_spectrum_single_electron} and simplify the resulting expression using the vector identity $[\rhat \times (\rhat \times \boldsymbol{a})]\cdot [\rhat \times (\rhat \times \boldsymbol{b})] = (\rhat \times \boldsymbol{a}) \cdot (\rhat \times \boldsymbol{b})$. We then obtain
\begin{widetext}
\begin{align}
   \frac{\partial^2 \SpectralEnergy_{\rm rad}}{\partial \nu \partial \Omega}(\rhat, \nu) &= \frac{q^2}{8 \pi^2 \eps_0 c} \Bigg\vert 2i\pi \nu \int_{t_{\nameRear}-\delta t}^{t_\nameTraj+\delta t} \rhat \times
    \left[ \fatbeta_p^-(t) e^{-i \Theta^-(t)}  - \fatbeta_p^+(t)  e^{-i \Theta^+(t)} \right] \dd t \nonumber \\
    &+ \left[ \frac{\rhat \times \fatbeta_p^-(t)}{1-\fatbeta_p^-(t)\cdot \rhat} e^{-i \Theta^-(t)}  - \frac{\rhat \times \fatbeta_p^+(t)}{1-\fatbeta_p^+(t) \cdot \rhat} e^{-i \Theta^+(t)} \right]_{t_{\nameRear} - \delta t}^{t_{\nameTraj}+\delta t}
    \Bigg\vert^2 \,,
    \label{equ:ctr:integration_by_parts_2}
\end{align}
\end{widetext}
where $\Theta^\pm (t) = 2 \pi \nu [t - \rhat \cdot \r_p^\pm (t)/c]$.
Since, from Eqs.~\eqref{equ:ctr:one_electron:acceleration_profile}, $\fatbeta_p(t_{\nameTraj} + \delta t) = \fatbeta_p(t_{\nameRear} - \delta t) = 0$, the boundary value terms vanish and the full intensity distribution radiated by the electron is therefore given by
\begin{align}
    \label{equ:ctr:full_radiation}
    &\frac{\partial^2 \SpectralEnergy_{\rm TSR}}{\partial \nu \partial \Omega}(\rhat, \nu) = \frac{q^2 \nu^2}{2 \eps_0 c} \nonumber \\
    &\times \left\vert \int_{t_\nameRear}^{t_\nameTraj} \rhat \times
    \left[\fatbeta_p^-(t) e^{-i \Theta^{-}(t)} 
    - \fatbeta_p^+(t)  e^{-i \Theta^{+}(t)} \right] \,\dd t \right\vert^2 \,,
\end{align}
where \textReferee{we have taken the limit $\delta t \to 0^+$}. The acronym TSR stands for transition and synchrotron radiation. Its coherent version (CTSR) in the case of a compact electron bunch will be addressed below.
\textReferee{Only this total radiation, which depends on the complete electron trajectory in vacuum, can be measured by a detector and thus has a tangible physical meaning. For didactic purposes, however, it makes sense to break it down into known radiation mechanisms, namely, transition radiation and synchrotron radiation, in order to show how destructively they interfere to produce it.}

The intensity distribution of transition radiation, $\partial^2 \SpectralEnergy_{\rm TR}/\partial \nu \partial \Omega$, is obtained from Eq.~\eqref{eq:frequency-angle_spectrum_single_electron} by integrating by parts over the intervals $t_{\nameRear}-\delta t < t < t_{\nameRear}$ and $t_{\nameTraj} < t < t_{\nameTraj} + \delta t$, and again taking $\delta t \to 0^+$. Only the boundary value term then remains, leading to a variant of the well-known Ginzburg formula \cite{Ginzburg:book:1979, *Ginzburg:ps:1982}
\begin{align}
    &\frac{\dpa^2 \SpectralEnergy_{\rm TR}}{\dpa \nu \dpa \Omega}( \rhat, \nu) = \frac{q^2}{8 \pi^2 \eps_0 c} \nonumber \\
    & \times \left\vert  \left[ \frac{\rhat \times \fatbeta_{\nameRear}^- }{1-\fatbeta_{\nameRear}^- \cdot \rhat} - \frac{\rhat \times \fatbeta_{\nameRear}^+}{1-\fatbeta_{\nameRear}^+ \cdot \rhat} \right] e^{-i \Theta_{\nameRear}} \right. \nonumber \\
    & \left. - \left[ \frac{\rhat \times \fatbeta_{\nameTraj}^- }{1-\fatbeta_{\nameTraj}^- \cdot \rhat} - \frac{\rhat \times \fatbeta_{\nameTraj}^+}{1-\fatbeta_{\nameTraj}^+ \cdot \rhat} \right] e^{-i \Theta_{\nameTraj}} \right\vert ^2 \,,
    \label{equ:ctr:one_electron:tr_spectrum}
\end{align}
where $\Theta_{\nameRear, \nameTraj} \equiv 2 \pi \nu (t_{\nameRear,\nameTraj} - \rhat \cdot \r_{\nameRear, \nameTraj}/c)$.

Similarly, the synchrotron component of the radiation is obtained by integrating Eq.~\eqref{eq:frequency-angle_spectrum_single_electron} by parts over the time interval $t_{\nameRear} < t < t_{\nameTraj}$:
\begin{align}
    &\frac{\partial^2 \SpectralEnergy_{\rm SR}}{\partial \nu \partial \Omega}(\rhat, \nu) = \frac{q^2}{8\pi^2 \eps_0 c} \nonumber \\
    &\times \Bigg\vert 2i \pi \nu \int_{t_{\nameRear}}^{t_\nameTraj} 
   \rhat \times \left[ \fatbeta_p^{-}(t) e^{-i \Theta^{-}(t)} - \fatbeta_p^{+}(t) e^{-i \Theta^{+}(t)}\right] \,\dd t \nonumber \\
    & + \left[ \frac{\rhat \times \fatbeta_{\nameTraj}^- }{1-\fatbeta_{\nameTraj}^- \cdot \rhat} - \frac{\rhat \times \fatbeta_{\nameTraj}^+}{1-\fatbeta_{\nameTraj}^+ \cdot \rhat} \right] e^{-i \Theta_{\nameTraj}}  \nonumber \\
    &  - \left[ \frac{\rhat \times \fatbeta_{\nameRear}^- }{1-\fatbeta_{\nameRear}^- \cdot \rhat} - \frac{\rhat \times \fatbeta_{\nameRear}^+}{1-\fatbeta_{\nameRear}^+ \cdot \rhat} \right] e^{-i \Theta_{\nameRear}}
    \Bigg\vert^2 \,,
    \label{equ:ctr:one_electron:sr_spectrum}
\end{align}

Of course, one recovers the full spectrum \eqref{equ:ctr:full_radiation} by summing the terms within the squared vertical bars in Eqs.~\eqref{equ:ctr:one_electron:tr_spectrum} and \eqref{equ:ctr:one_electron:sr_spectrum}. The above expression is interesting in showing that whatever the evolution of $\fatbeta^\pm(t)$ during $t_{\nameRear} < t < t_{\nameTraj}$, the synchrotron radiation generates a spectrum increasingly resembling that due to transition radiation when $t_{\nameTraj} - t_{\nameRear} \equiv \Delta t_r \to 0$. This result will be demonstrated numerically in the next sections. 

In the following, for simplicity, the electric sheath field acting on the electron will be assumed uniform, stationary and parallel to the target surface normal, $\E_0=E_0 \hat{\boldsymbol{z}}$ (with $E_0>0$). The electron trajectory can be easily calculated using the Hamiltonian
\begin{equation}
    \label{equ:ctr:one_electron:hamiltonian}
    \mathcal{H} = m_e c^2 \gamma(t) - e\SheathPotential(z(t))\,,
\end{equation}
where $\gamma \equiv \sqrt{1+p^2/(m_e c)^2}$ is the electron Lorentz factor and $\SheathPotential = -E_0 z$ the electrostatic potential. 

The electron crosses the target backside ($z=0$) at time $t=t_{\nameRear}$ and transverse position $\r_{\nameRear, \perp}$ (neglecting the $\delta t$ interval) with momentum $\p_{\nameRear} = p_{\nameRear} (\sin \psi \cos \varphi, \sin \psi \sin \varphi, \cos \psi)$, where $\psi$ and $\varphi$ correspond to the polar and azimuthal angles, respectively. Since $\SheathPotential$ only depends on $z$, $\mathcal{H}$ is a constant of motion. Moreover, $\gamma(t_{\nameTraj}) = \gamma(t_{\nameRear}) \equiv \gamma_{\nameRear}$ and $\boldsymbol{p}_\perp (t) = \boldsymbol{p}_{\nameRear, \perp}$ (the subscript $_\perp$ refers to the surface normal plane).

Introducing the normalized momentum $\boldsymbol{u} \equiv \boldsymbol{p}/m_e c \equiv \gamma \fatbeta$, normalized position $\bar{\r} \equiv e E_0 \r /m_e c^2$, relative position $\fatxibar = \bar{\r} - \bar{\r}_{\nameRear,\perp}$, normalized time $\tbar \equiv e E_0 t/m_e c$ and relative time $\taubar = \tbar - \tbar_{\nameRear}$, one obtains
\begin{align}
    &\gamma(\taubar) = \sqrt{\left(u_{\nameRear,z} - \taubar\,\right)^2 + \gamma_{\perp}^2} \,, \label{eq:gamma} \\[4pt]
    &\beta_z(\taubar) = \left(u_{\nameRear,z} - \taubar\,\right)/\gamma(\taubar) \,, \label{eq:beta_z} \\[4pt]
    &\fatbeta_\perp(\taubar) = \boldsymbol{u}_{\nameRear,\perp}/\gamma(\taubar) \,, \label{eq:beta_perp} \\[4pt]
    &\xibar_z(\taubar) = \gamma_{\nameRear} - \gamma(\taubar) \,, \label{eq:z} \\[4pt]
    &\fatxibar_\perp (\taubar) = \frac{\boldsymbol{u}_{\nameRear, \perp}}{2} 
    \ln \left(\frac{ \left(\gamma_{\nameRear} + u_{\nameRear,z} \right) \left[ \gamma(\taubar) - u_{\nameRear,z} + \taubar \right] }{\left(\gamma_{\nameRear} - u_{\nameRear,z} \right) \left[\gamma(\taubar) + u_{\nameRear,z} - \taubar \right]} \right) \,, \label{eq:r_perp}
\end{align}
for $0 < \taubar < \DeltaRetBar$, where $\DeltaRetBar \equiv \tbar_{\nameTraj} - \tbar_{\nameRear} = 2 u_{\nameRear,z}$ is the (normalized) time spent by the electron in vacuum, and $\gamma_\perp = \sqrt{1+u_{\nameRear, \perp}^2}$.

The full (TSR) and synchrotron-only (SR) radiation spectra are computed by substituting the above expressions into Eq.~\eqref{equ:ctr:full_radiation} and Eq.~\eqref{equ:ctr:one_electron:sr_spectrum}, respectively. Furthermore, given our choice of $\E_0$, we have the relations $\fatbeta_{\nameTraj}^{\pm} = \fatbeta_{\nameRear}^{\mp}$ and $\beta_{\nameTraj, z}^{\pm}=-\beta_{\nameRear, z}^{\pm}$, allowing the TR spectrum \eqref{equ:ctr:one_electron:tr_spectrum} to be recast as
\begin{align}
    &\frac{\dpa^2 \SpectralEnergy_{\rm TR}}{\dpa \nu \dpa \Omega}( \rhat, \nu) = \frac{q^2}{2 \pi^2 \eps_0 c} \cos^2 \left( \frac{\Theta_{\nameRear} - \Theta_{\nameTraj}}{2} \right) \nonumber \\
    &\times \left\vert \frac{\rhat \times \fatbeta_{\nameRear}^{-}}{1 - \fatbeta_{\nameRear}^{-} \cdot \rhat} - \frac{\rhat \times \fatbeta_{\nameTraj}^{-}}{1 - \fatbeta_{\nameTraj}^{-} \cdot \rhat} \right\vert^2 \,.
    \label{equ:ctr:one_electron:tr_spectrum2}
\end{align}

%
%
\subsection{Full radiation from an energy-angle-distributed electron beam}
\label{subsec:ctr:electron_beam}

The frequency-angle spectrum radiated by an electron bunch exiting a perfect conductor and experiencing a uniform electric field can be readily evaluated by summing Eq.~\eqref{equ:ctr:full_radiation} over $N_h \gg 1$ particles (labeled by the subscript $l$) and taking an ensemble average:
\begin{equation}
\label{equ:ctr:electron_beam:radiated_energy_sum_l}
    \frac{\partial^2 \ICTSR}{\partial \nu \partial \Omega}(\rhat, \nu) =  \left\langle \frac{q^2 \nu^2}{2 \eps_0 c} \bigg\vert \sum_{l=1}^{N_h} \big[ \A_l^\nameRealParticle - \A_l^\nameImageParticle \vphantom{\int} \big] (\rhat, \nu) \bigg\vert^2 \right\rangle\,,
\end{equation}
where
\begin{equation}
    \label{equ:ctr:electron_beam:definition_of_A}
    \A_l^\pm (\rhat, \nu) = \int_{t_{\nameRear, l} }^{t_{\nameTraj,l}} 
    \left[\rhat \times \fatbeta_{p,l}^\pm (t) \right] e^{-i \Theta_l^\pm(t)} \,\dd t \,.
\end{equation}
We recall that the acronym CTSR stands for \emph{coherent} transition and synchrotron radiation.

In the realistic case of a finite-size, energy-angle distributed electron bunch, the full radiation pattern is determined by the relative phase shifts between the fields emitted along the particle trajectories in vacuum. In the following, each electron trajectory is taken to originate from the \emph{front side} ($z=-d$) of the target, where (leaving out the subscript $l$) it is parameterized by the injection time $t_{\nameFront}$, initial transverse position $\r_{\nameFront, \perp}$ and initial normalized momentum $\u_{\nameFront} = \u_{\nameRear} = \p_{\nameRear}/m_e c$ of the electron. After traveling ballistically through the target, each electron reaches its rear side ($z=0$) at time $t_{\nameRear} = t_{\nameFront} + d/ c \beta_{\nameFront, z}$ and transverse position $\r_{\nameRear, \perp} = \r_{\nameFront, \perp} + \fatbeta_{\nameFront, \perp} d/\beta_{\nameFront, z}$.

In order to pass to the continuous limit, we introduce $\FluxProbabilityFunction(\r,t,\u)$, the flux probability function characterizing the electron beam at the front side of the target, normalized as 
\begin{equation}
    \label{equ:ctr:electron_beam:flux_probability}
    \iiint \FluxProbabilityFunction(\r_{\nameFront,\perp},t_{\nameFront},\u_{\nameFront}) \, \dd^2 \r_{\nameFront, \perp}\,\dd t_{\nameFront} \, \dd^3 \u_{\nameFront}  = 1\,.
\end{equation}

This probability function is taken in the form of a separable function, $\FluxProbabilityFunction(\r_{\nameFront, \perp},t_{\nameFront},\u_{\nameFront}) = f(\r_{\nameFront,\perp}) h(t_{\nameFront}) g(\u_{\nameFront}) $, where the distribution functions $f$, $h$ and $g$ will be specified in \seccite{\ref{subsec:ctr:parameters}}.
The discrete sum in Eq.~\eqref{equ:ctr:electron_beam:radiated_energy_sum_l} can be replaced by integrals over time, transverse position and momentum (see \appcite{\ref{sec:app:discrete_to_continuous_description}}):
\begin{align}
    \label{equ:ctr:electron_beam:radiated_energy_integral}
    &\frac{\partial^2 \ICTSR}{\partial \nu \partial \Omega}(\rhat, \nu) =  \frac{q^2 \nu^2}{2 \eps_0 c} N_h^2 \Big\vert \iiint f(\r_{\nameFront,\perp}) h(t_{\rm f}) g(\u_{\nameFront}) \nonumber \\
    & \times \big[ \A^\nameRealParticle(\rhat,\nu) - \A^\nameImageParticle(\rhat,\nu) \big] \,\dd^2 \r_{\nameFront,\perp} \,\dd t_{\nameFront} \,\dd^3 \u_{\nameFront} \Big \vert^2\,.
\end{align}

This formula can further be simplified by rewriting the phase term in Eq.~\eqref{equ:ctr:electron_beam:definition_of_A} as
\begin{align}
       \Theta^\pm (t) &= 2 \pi \nu \left[ t - \rhat \cdot \r^\pm (t)/c \right] \nonumber \\[4pt]
       &= 2 \pi \nubar \left[ (\taubar + \tbar_{\nameRear}) - \rhat \cdot (\fatxibar^\pm(\taubar) + \bar{\r}_{\nameRear,\perp}) \right] \nonumber \\[4pt]
       &= \Theta_{\nameRear} + 2 \pi \nubar \left[ \taubar - \rhat \cdot \fatxibar^\pm(\tbar) \right]\,,
\end{align}
where $\nubar \equiv \nu m_e c / e E_0$ is the normalized frequency and where
\begin{align}
    \label{equ:ctr:one_electron:phase_eitheta}
    \Theta_{\nameRear} &= 2 \pi \nubar \left[ \tbar_{\nameRear} - \rhat \cdot \bar{\r}_{\nameRear,\perp} \right] \nonumber \\
    &= 2 \pi \nubar \left(\tbar_{\nameFront} - \rhat \cdot \bar{\r}_{\nameFront}\right) + 2 \pi \nubar \bar{d} \left(1 - \rhat \cdot \fatbeta_{\nameFront,\perp}\right) / \beta_{\nameFront, z}
\end{align}
characterizes the propagation of the electron through the foil. Performing the change of variable $t \mapsto \taubar$ and factoring out $\Theta_{\nameRear}$ from Eq.~\eqref{equ:ctr:electron_beam:definition_of_A} yields

\begin{align}
    &\frac{\partial^2 \ICTSR}{\partial \nu \partial \Omega}(\rhat, \nu) = 
    \frac{q^2 \nubar^2}{2 \eps_0 c} N_h^2 \left\vert \int g(\u) F(\rhat,\nu,\u) \right. \nonumber \\
    &\times \left. \Big[ \widebar{\A}^\nameRealParticle (\rhat,\nu) - \widebar{\A}^\nameImageParticle (\rhat,\nu) \vphantom{\int} \Big] \, \dd^3 \u \right\vert^2\,,
    \label{equ:ctr:electron_beam:radiated_energy_Fourier_integrated}
\end{align}
with $F(\rhat,\nu,\u)$ the form factor of the electron beam defined by
\begin{equation}
    \label{equ:ctr:electron_beam:define_the_form_factor}
    F(\rhat,\nu,\u) = \iint f(\r_{\nameFront,\perp}) h(t_{\rm f}) e^{- i \Theta_{\nameRear}} \dd t_{\rm f}\,\dd^2 \r_{\nameFront,\perp} \,,
\end{equation}
and where
\begin{equation}
    \label{equ:ctr:electron_beam:definition_of_A_rel}
    \widebar{\A}^\pm (\rhat,\nu) = \int_{0}^{\DeltaRetBar} 
    \left[\rhat \times \fatbeta^\pm (\taubar) \right] e^{-2i\pi \nubar \left[ \taubar - \rhat \cdot \fatxibar^\pm (\taubar) \right]} \,\dd \taubar \,.
\end{equation}

Equations~\eqref{equ:ctr:electron_beam:radiated_energy_Fourier_integrated}-\eqref{equ:ctr:electron_beam:definition_of_A_rel}, supplemented with Eqs.~\eqref{eq:gamma}-\eqref{eq:r_perp}, give the energy-angle spectrum of radiation from an electron beam exiting a perfect conductor and being reflected by a stationary and homogeneous electric sheath field. This spectrum depends not only on the distribution functions $f(\r_\perp)$, $h(t)$ and $g(\u)$ characterizing the beam source at the target front side, but also on the finite target thickness, which determines the longitudinal and transverse spreading of the beam after its passage through the foil. 

As explained in \seccite{\ref{subsec:ctr:one_electron_trajectory}}, the respective contributions of CTR ($\partial^2 \SpectralEnergy_{\rm CTR}/\partial \nu \partial \Omega$) and CSR ($\partial^2 \SpectralEnergy_{\rm CSR}/\partial \nu \partial \Omega$) to the total spectrum can be distinguished in the time integral involved in $\bar{\A}^\nameMultipleParticles(\r,\nu)$ [Eq. (\ref{equ:ctr:electron_beam:definition_of_A_rel})], namely,
\begin{widetext}
\begin{align}
    \frac{\partial^2 \SpectralEnergy_{\rm CTR}}{\partial \nu \partial \Omega}(\rhat, \nu) & = \frac{N_h^2q^2}{2 \pi^2 \eps_0 c } \left\vert \int g(\u) F(\theta, \nu,\u) \AInterfaces \, \dd^3 \u \right\vert^2 \,, 
    \label{equ:ctr:electron_beam:radiated_energy_CTR}\\
    \frac{\partial^2 \SpectralEnergy_{\rm CSR}}{\partial \nu \partial \Omega}(\rhat, \nu) & = \frac{N_h^2q^2}{2 \pi^2 \eps_0 c} \left\vert \int g(\u) F(\theta, \nu,\u) \Biggl\{ \vphantom{\int} i \pi \nubar \Bigl[ \widebar{\A}^\nameRealParticle (\rhat,\nu) - \widebar{\A}^\nameImageParticle (\rhat,\nu) \Bigr] - \AInterfaces \Biggr\} \, \dd^3 \u \right\vert^2 \,,
    \label{equ:ctr:electron_beam:radiated_energy_CSR}
\end{align}
\end{widetext}
where $\AInterfaces$ characterizes the full transition radiation of a particle in the vicinity of the perfect conductor [stages $\rm (iii)$ and $\rm (v)$ in \figcite{\ref{fig:ctr:illustration:target_thin_foil}}] and is defined by:
\begin{multline}
    \label{equ:ctr:electron_beam:A_interface}
    \AInterfaces = \cos \left( \frac{\Theta_{\nameRear} - \Theta_{\nameTraj}}{2} \right) \\
     \times \left[ \frac{\rhat \times \fatbeta_{\nameRear}^{-}}{1 - \fatbeta_{\nameRear}^{-} \cdot \rhat} - \frac{\rhat \times \fatbeta_{\nameTraj}^{-}}{1 - \fatbeta_{\nameTraj}^{-} \cdot \rhat} \right] e^{-i \Theta_{\nameRear}}\,,
\end{multline}
where $\Theta_{\nameTraj} = \Theta_{\nameRear} + 2 \pi \nubar \left[ \widebar{\Delta t_{\nameTraj}} -\rhat \cdot \fatxibar \left( \widebar{\Delta t_{\nameTraj}} \right) \right]$.

The numerical evaluation of Eqs.~\eqref{equ:ctr:electron_beam:radiated_energy_Fourier_integrated}, \eqref{equ:ctr:electron_beam:radiated_energy_CTR} and \eqref{equ:ctr:electron_beam:radiated_energy_CSR} is challenging because of the four nested integrals involved, or even seven if one wishes to compute the total radiated energy (integrated over observation angles $(\theta, \Psi)$ and frequencies $\nu$). To simplify, we will assume hereafter that the system is axisymmetric with respect to the target normal ($\zhat$ axis). The integration over $(\rhat,\nu)$ then reduces to an integration over $(\theta,\nu)$ due to invariance of $\partial^2 \SpectralEnergy / \partial \nu \partial \Omega$ over $\Psi$. In this work, the calculation over the angle of observation, frequency and momentum is parallelized on multiple GPUs, using a 2D kernel $(\theta \times \nu, u \times \psi \times \varphi)$ to efficiently distribute the workload. The innermost time integration [Eq.~\eqref{equ:ctr:electron_beam:definition_of_A_rel}] is performed over 1000 time steps through a type-2 nonuniform Fourier transform, as defined in Refs.~\onlinecite{Dutt:siam:1993, Greengard:siam:2004, Ruiz:siam:2018}. Numerical integrals over $(u,\psi,\varphi)$ are computed using $512 \times 128 \times 64$ points for each $(\nu,\theta)$ pair. Radiated spectra are discretized over $160 \times 90$ points in $(\nu, \theta)$ space, amounting to a total of $\sim 6 \times 10^{10}$ nonuniform Fourier transforms. This makes parametric scans quite computationally expensive despite the GPU parallelization.

\subsection{Model parameters}
\label{subsec:ctr:parameters}

To solve the previous formulas, the distribution functions defining the (axisymmetric) electron beam source need to be specified. The spatial and temporal profiles of the beam are assumed to be Gaussian,
\begin{align}
    \label{eq:f}
    f(r) &= \frac{4 \ln 2}{\pi w_L^2} e^{-4 \ln 2\,(r / w_L)^2} \,, \\
    \label{eq:h}
    h(t) &= \frac{2 \sqrt{\ln 2}}{\sqrt{\pi} \tau_L} e^{-4 \ln 2\,(t /\tau_L)^2} \,,
\end{align}
where the full-width-at-half-maximum diameter ($w_L$) and duration ($\tau_L$) are taken equal to those characterizing the intensity profile of the laser drive. The above profiles lead to the form factor
\begin{align}
    \label{equ:ctr:electron_beam:form_factor_calculated}
    F(\theta, \nu,\u) = & \exp\Bigl( -(\pi^2/4 \ln 2) \nu^2 \left[ \tau_L^2 + (w_L/c)^2 \sin^2 \theta \right] \Bigr) \nonumber \\
    \times & \exp\Bigl( -2i \pi \nu d \left( 1 - \rhat_\perp \cdot \fatbeta_{\perp} \right) / c \beta_z \Bigr) \,.
\end{align}

We further assume that the momentum distribution function of the beam is a separable function of absolute momentum $u$ and direction angles $(\psi, \varphi)$, i.e., $g(\u) \dd^3 \u = g_u(u) g_\psi(\psi) g_\varphi(\varphi) u^2 \sin \psi\, \dd u\,\dd \psi \,\dd \varphi$. Inspired by Ref.~\onlinecite{Schroeder:pre:2004}, we use
\begin{align}
    g_u(u) &= \frac{1}{2 \Delta u^3} \exp(-u/\Delta u)\,,\\
    g_\psi(\psi) &= \frac{1}{\Delta \psi^2} \frac{e^{-\frac{1}{2}\left( \sin \psi / \Delta \psi \right)^2 }}{1 - e^{-1 / 2\Delta \psi^2}} \cos \psi \,,\\
    g_\varphi(\varphi) &= 1/2\pi\,,
\end{align}
where $\Delta u$ and $\Delta \psi \equiv \sin \Psi_h$ represent characteristic spreads in momentum and polar angle, respectively. The above distributions are normalized such that
\begin{equation}
    \label{equ:ctr:electron_beam_normalization_of_momentum_distribution_functions}
    \int_0^\infty g_u(u) \,u^2\,\dd u = \int_0^{\pi/2}  g_\psi(\psi)\,\sin \psi\,\dd \psi = 1\,.
\end{equation}

The momentum spread $\Delta u$ is chosen so that the average Lorentz factor
\begin{equation}
    \label{equ:ctr:gamma_h_average:definition}
    \gammaAvg = \int_0^\infty g_u(u) u^2 \sqrt{u^2 + 1} \,\dd u \sim \sqrt{ 9\Delta u^2 + 1}
\end{equation}
equals the ponderomotive potential of the (linearly polarized) laser pulse, $\gamma_L = \sqrt{1+a_L^2/2}$, where $a_L$ is the dimensionless laser field strength \cite{Macchi:book:2013}. 

The number of hot electrons carried by the beam is estimated from
\begin{equation}
    \label{equ:ctr:number_of_electrons:definition}
    N_h \simeq \frac{\eta_h \Energy_L}{m_e c^2 \langle \gamma_h - 1 \rangle} = \frac{\Energy_{\rm beam}}{\langle \Energy_h \rangle}\,,
\end{equation}
where $\Energy_L = I_L (w_L^2 \tau_L/8) (\pi/\ln2)^{3/2}$ is the laser pulse energy, $\langle \Energy_h \rangle = m_e c^2 \langle \gamma_h-1\rangle$ the mean electron kinetic energy, $\eta_h$ the laser-to-hot-electron energy conversion efficiency, and $\Energy_{\rm beam}= \eta_h \Energy_{L}$ the total kinetic beam energy. Based on experimental measurements~\cite{Nilson:pop:2008, Carroll:njp:2010}, a constant value $\eta_h = 0.2$ will be assumed in the following.

Introducing $n_{\rm hr0}$ as the hot-electron density at the target backside, the strength of the sheath field is expected to be~\cite{Fill:pop:2001, Mora:prl:2003}
\begin{equation}
    \label{equ:E0_field}
    E_0 \simeq \alpha_{E_0} \sqrt{\frac{n_{\rm hr0} \langle \Energy_h \rangle}{\eps_0}} \,,
\end{equation}
where $\alpha_{E_0} \le 1$ is a scaling factor expressing that the effective average field strength experienced by the hot electrons in vacuum is lower than its maximum value at $z=0$, $E_{\rm max} = \sqrt{n_{\rm hr0} \langle \Energy_h \rangle/\eps_0}$. We will take $\alpha_{E_0}=0.5$ as a fiducial parameter. Moreover, $n_{\rm hr0}$ can be related to the hot-electron density at the front side, $n_{\rm hf0}$, defined by
\begin{equation}
   n_{\rm hf0} \simeq  \frac{\eta_h I_L}{m_ec^3 \langle \gamma_h - 1 \rangle} \,.
\end{equation}
through
\begin{equation}
    \label{equ:ctr:nhr0}
    n_{\rm hr0} = n_{\rm hf0} \left(\frac{w_L}{w_h(d)} \right)^2\,.
\end{equation}
The transverse size of the beam, $w_h(z)$, after a ballistic propagation of a depth $z$ through the target is estimated~\cite{Debayle:pre:2010} as
\begin{equation}
    \label{equ:ctr:whz}
    w_h(z) \simeq w_L \sqrt{1 + \left( \frac{2 z \tan \Psi_h}{w_L}\right)^2} \,.
\end{equation}
The sheath field $E_0$ is obtained from Eqs.~\eqref{equ:E0_field}-\eqref{equ:ctr:whz}.

\section{Phenomenology of CTSR and parametric dependencies}
\label{sec:ctr:model_results}

\subsection{Interplay of transition and synchrotron radiations}
\label{subsec:ctr:interplay}

\begin{figure*}
 \centering
 \includegraphics[width=0.99\linewidth]{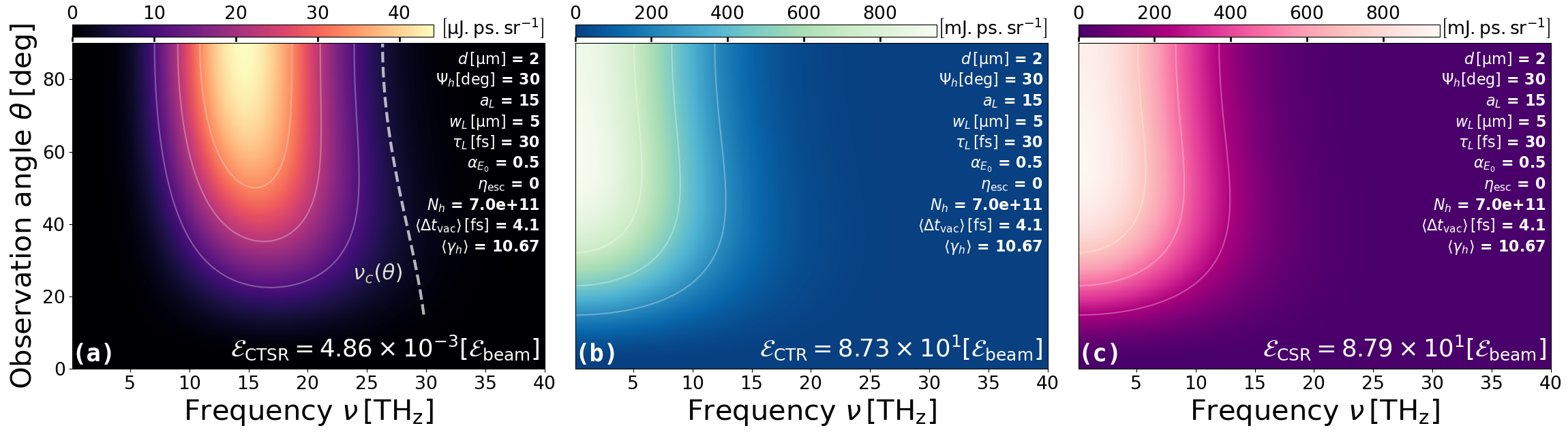}
 \caption{(a) Frequency-angle spectrum $\partial^2 \ICTSR/\partial \nu \partial \Omega$ (in $\rm \mu J\,ps\,sr^{-1}$ units) radiated by hot electrons in a conducting  foil of $d=2\,\rm \mu m$ thickness, exposed to a $a_L=15$, $w_L=5\,\rm \mu m$, $\tau_L=30\,\rm fs$, $\Energy_L = 2.8\,\rm J$ laser pulse. The laser-to-hot-electron conversion efficiency is set to $\eta_h = 0.2$, leading to an electron beam energy of $\Energy_{\rm beam} = 0.56\,\rm J$. The white dashed line plots Eq.~\eqref{equ:ctr:cutoff_frequency} with $\eta_{\rm coh} = 3 \times 10^{-3}$. Panels (b) and (c) show the spectra (in $\rm mJ\,ps\,sr^{-1}$ units) due to solely CTR and CSR, respectively.}
 \label{fig:ctr:electron_beam:radiat_mech_comp}
\end{figure*}

As a first illustration of our modeling, we consider the case of a $2\,\rm \mu m$ thick target exposed to an intense laser pulse characterized by $a_L = 15$, $\lambda_L=1\,\rm \mu m$, $\tau_L = 30\,\rm fs$ and $w_L = 5\,\rm \mu m$. These parameters yield a peak pulse intensity $I_L=3.1\times 10^{20}\,\rm W\,cm^{-2}$ and pulse energy $\Energy_L  = 2.8\,\rm J$. The beam divergence is taken to be $\Psi_h = 30\deg$. For these parameters (together with $\eta_h=0.2$), one obtains $\Energy_{\rm beam} \simeq 0.56\,\rm J$, $\Energy_h \simeq 4.9\,\rm MeV$, $N_h \simeq 7.03\times 10^{11}$ and $E_0 \simeq 6.93\times 10^{12}\,\rm V\,m^{-1}$.

Figure~\ref{fig:ctr:electron_beam:radiat_mech_comp}(a) shows the angular distribution of the full energy spectrum (in $\rm \mu J\,ps \,sr^{-1}$ units) for $0.1 \le \nu \le 40\,\rm THz$, while Figs.~\ref{fig:ctr:electron_beam:radiat_mech_comp}(b) and (c) display the spectra (in $\rm mJ\,ps \,sr^{-1}$ units) due to only CTR and CSR, respectively. One can see that the full spectrum is peaked around $\nu \simeq 10-25\,\rm THz$ and $\theta \simeq 60-90\deg$.
By contrast, CTR and CSR give rise separately to almost identical -- but much more intense -- THz spectra concentrated at lower frequencies and angles.

The latter result can be readily understood upon noting that the first (second) half of the electron trajectory in vacuum, before (resp. after) its turning point, can be viewed, given its femtosecond timescale ($t \sim m_e \gamma_\nameRear c/eE_0 \sim 1\,\rm fs$), as a sudden deceleration along $\zhat$ (resp. acceleration towards $-\zhat$) with respect to THz-range synchrotron radiation. Since, by contrast, transition radiation at the exit (return) time corresponds to a sudden apparent acceleration (resp. deceleration), the two radiation mechanisms generate very similar THz fields but of opposite polarity and so almost exactly cancel out, especially at low frequencies ($\lesssim 10\,\rm THz$ here) and angles ($\lesssim 40\deg$). Mathematically speaking, this description merely amounts to taking $\DeltaRet \to 0$ in Eq.~\eqref{equ:ctr:electron_beam:definition_of_A_rel} so that Eq.~\eqref{equ:ctr:electron_beam:radiated_energy_CSR} converges to Eq.~\eqref{equ:ctr:electron_beam:radiated_energy_CTR}.
It is to be noted that the average time spent by the electrons in vacuum, $\langle \DeltaRet \rangle$, can be exactly solved as
\begin{align}
    \label{equ:ctr:electron_beam:average_delta_vac:definition}
    \langle \DeltaRetBar \rangle &= \iint g(u,\psi) \DeltaRetBar u^2 \sin \psi \,\dd^2 u \dd \psi\\
    \label{equ:ctr:electron_beam:average_delta_vac:computation}
    &= -3 \Delta u \left( 1 + \coth  \frac{X_\psi^2}{2} \right) \left( X_\psi^{-1} D\left( X_\psi \right) -1 \right)\,,
\end{align}
where $D(x) = e^{-x^{2}}\int_0^x e^{t^2}\,\dd t$ is the Dawson function, $X_\psi \equiv 1/\sqrt{2} \Delta \psi$ and $\DeltaRetBar = 2 u_z = 2 u \cos \psi$. For the parameters of Fig.~\ref{fig:ctr:electron_beam:radiat_mech_comp}, one finds $\langle \DeltaRet \rangle \simeq 4\,\rm fs$.

Such a strong interplay of CTR and CSR therefore precludes their separate treatment, otherwise the low-frequency part of the spectrum would be inaccurately modeled and, worse, the radiated energy would be unphysically overestimated. Indeed, while the integration of $\partial^2 \ICTSR/\partial \nu \partial \Omega$ over $0.1 < \nu < 40\,\rm THz$ and $\Omega$ gives a total radiated energy of $\Energy_{\rm CTCR} \simeq 4.86\times 10^{-3}\, \Energy_{\rm beam} \simeq 2.7\,\rm mJ$, the CTR and CSR spectra, both scaling as $N_h^2$, contain alone an energy much exceeding the total beam energy, i.e., $\Energy_{\rm CTR} \simeq \Energy_{\rm CSR} \simeq 88\,\Energy_{\rm beam}$.

Besides the high-pass filtering effect caused by the ultrashort reflection timescale of the electrons in vacuum, the form factor $F(\theta, \nu, \u)$ that characterizes the spatiotemporal coherence of the entire beam acts as a low-pass filter. The high-frequency shape of the spectrum can be approximately captured by the attenuation factor $\eta_{\rm coh}$ defined by
\begin{align}
    \label{equ:ctr:electron_beam:definition_attenuation_factor}
    \nonumber \eta_{\rm coh}(\theta, \nu) &= \left| \int g(\u) F(\theta, \nu, \u) \dd^3 \u \right|^2 \\
    &= C_{\rm coh}\,e^{-(\pi^2/2 \ln 2) \nubar^2 \left(\taubar_L^2 + \bar{w}_L^2 \sin^2 \theta \right)} \,,
\end{align}
where
\begin{equation}
    \label{equ:ctr:electron_beam:Contrast}
    C_{\rm coh} = \left| \int g(\u) e^{-2i \pi \nubar \bar{d} \left[ 1 - \rhat \cdot \fatbeta_{\perp} \right] / \beta_{z}} \dd^3 \u \right|^2
\end{equation}
quantifies the coherence of the beam after its propagation through the target of thickness $d$. If the target is sufficiently thin, the propagation effects can be neglected and since $\int g(\u) \dd^3 \u = 1$, one has $C_{\rm coh} \simeq 1$. In this case, the contour lines of the low-pass filter can be computed as a function of $\theta$ and $\nu$ through
\begin{equation}
    \label{equ:ctr:electron_beam:attenuation_factor_simplified}
   \eta_{\rm coh}(\theta, \nu) \simeq e^{-(\pi^2/2 \ln 2) \nubar^2 \left(\taubar_L^2 + \bar{w}_L^2 \sin^2 \theta \right)}\,.
\end{equation}
As seen in Fig.~\ref{fig:ctr:electron_beam:radiat_mech_comp}(a), the radiated spectrum assumes the shape of a band located, for our parameters, between $\nu \simeq 10\,\rm THz$ and the approximate upper bound 
\begin{equation}
    \label{equ:ctr:cutoff_frequency}
    \nubar_c(\theta) \simeq \sqrt{\frac{- 2 \ln 2 \ln \eta_{\rm coh}}{\pi^2 \left( \taubar_L^2 + \bar{w}_L^2 \sin^2 \theta\right)} }\,.
\end{equation}
This expression is plotted as a white dashed line in Fig.~\ref{fig:ctr:electron_beam:radiat_mech_comp}(a) for $\eta_{\rm coh}(\theta, \nubar_c) = 3 \times 10^{-3}$, leading to $25 \le \nu_c(\theta) \le 30\,\rm THz$.

\subsection{Influence of the sheath field}
\label{subsec:ctr:return_time}

Equation~\eqref{equ:E0_field} represents a crude expression of the sheath field strength which, in reality, varies both with space and time \cite{Fill:pop:2001, Mora:prl:2003}. To examine the dependency of the THz radiation spectrum on $E_0$, we perform a parametric study over $\alpha_{E_0}$.

Increasing the sheath field strength shortens the electron excursion time in vacuum: the CTR and CSR fields then compensate each other even better so that the energy radiated by CTSR is further reduced. Conversely, decreasing $E_0$ allows the particles to propagate over larger distances and radiate more efficiently. 

\begin{figure}[b!]
    \centering
    \includegraphics[width=0.95\linewidth]{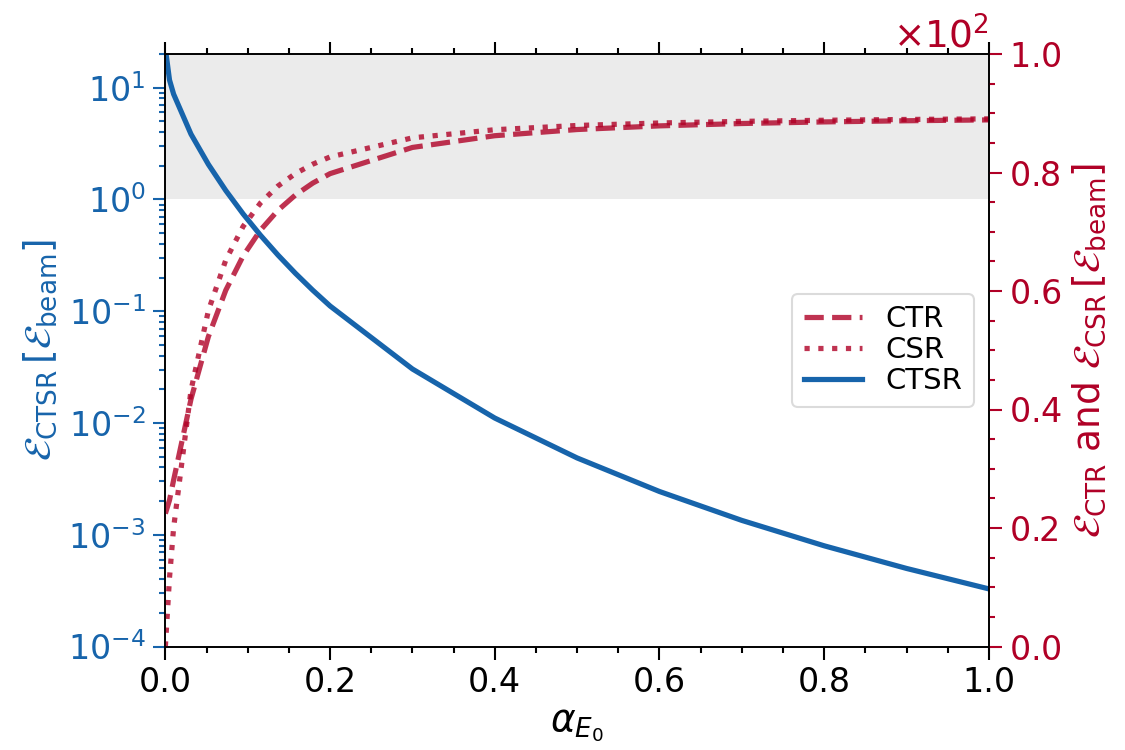}
    \caption{Radiated THz energy (integrated over solid angles and $0.1-40\,\rm THz$ frequencies) as a function of the sheath-field parameter $\alpha_{E_0}$, as defined by Eq.~\eqref{equ:E0_field}. The blue solid curve, in $\rm log_{10}$ scale, represents the full radiated energy while the red dashed and dotted curves plot the energy yields of CTR and CSR, respectively. The laser-plasma parameters are those used in Fig.~\ref{fig:ctr:electron_beam:radiat_mech_comp} ($a_L=15$).
    The reference case $\alpha_{E_0} = 0.5$ corresponds to a field $E_0 \simeq 6.9\times 10^{12}\,\rm V\,m^{-1}$.
    }
    \label{fig:ctr:electron_beam:parametric_studies:influence_of_E0}
\end{figure}

Figure~\ref{fig:ctr:electron_beam:parametric_studies:influence_of_E0} shows the variation in the radiated energy (integrated over solid angles and the $0.1-40$ THz frequency range) with $\alpha_{E_0} \le 1$, for the same laser-plasma parameters as in Fig.~\ref{fig:ctr:electron_beam:radiat_mech_comp}. 
The energies radiated through CTR and CSR, taken separately, decrease when lowering $\alpha_{E_0}$ whilst the total radiated energy rapidly rises (by $\sim 10\times$ from $\alpha_{E_0} = 1$ to $\alpha_{E_0}=0.5$, and by $\sim 20\times$ from $\alpha_{E_0} = 0.5$ to $\alpha_{E_0}=0.2$). As expected, CSR tends to vanish when $\alpha_{E_0} \to 0$ so that CTR then accounts for all of the radiated energy. Yet, the electrons then only radiate coherently while exiting the target because they are greatly spread out upon re-entering the foil. When $\alpha_{E_0} \to 1$, by contrast, they are reflected so rapidly into the target that they remain packed enough to emit two almost coincident CTR bursts [stages (iii) and (v) in Fig.~\ref{fig:ctr:illustration:target_thin_foil}] of same polarity (the first from the accelerating real particles, the second from the decelerating image particles), the total energy of which is four times that of the initial CTR burst.

Of course, $\alpha_{E_0}$ is just a fitting parameter, introduced for simplicity to capture the overall effect of the highly nonstationary sheath field \cite{Fill:pop:2001, Ridgers:pre:2011}. Precautions must therefore be taken when analyzing the radiated energy for very weak sheath fields. Notably, below $\alpha_{E_0} \simeq 0.2$, the hot electrons spend so much time ($\gtrsim 10\, \rm fs$) in vacuum before returning to the target that a sizable ($>0.1$) fraction of the beam energy is predicted to be radiated away through CTSR. Even worse, below $\alpha_{E_0} \simeq 0.1$, the total emitted energy approaches that potentially radiated by all accelerated electrons through CTR alone, entailing non-physically high ($>\Energy_{\rm beam}$) radiated energies. A proper treatment of the problem for such weak sheath fields would therefore require a self-consistent description of the backreaction of radiation on the electron dynamics, as is done, \emph{e.g.}, in accelerator physics \cite{Stupakov:book:2018}.
This is a challenging task, well exceeding the scope of the present semianalytical model.

\subsection{Contribution of escaping ballistic electrons}
\label{subsec:ctr:escaping_electrons}

\begin{figure}
    \centering
    \includegraphics[width=0.99\linewidth]{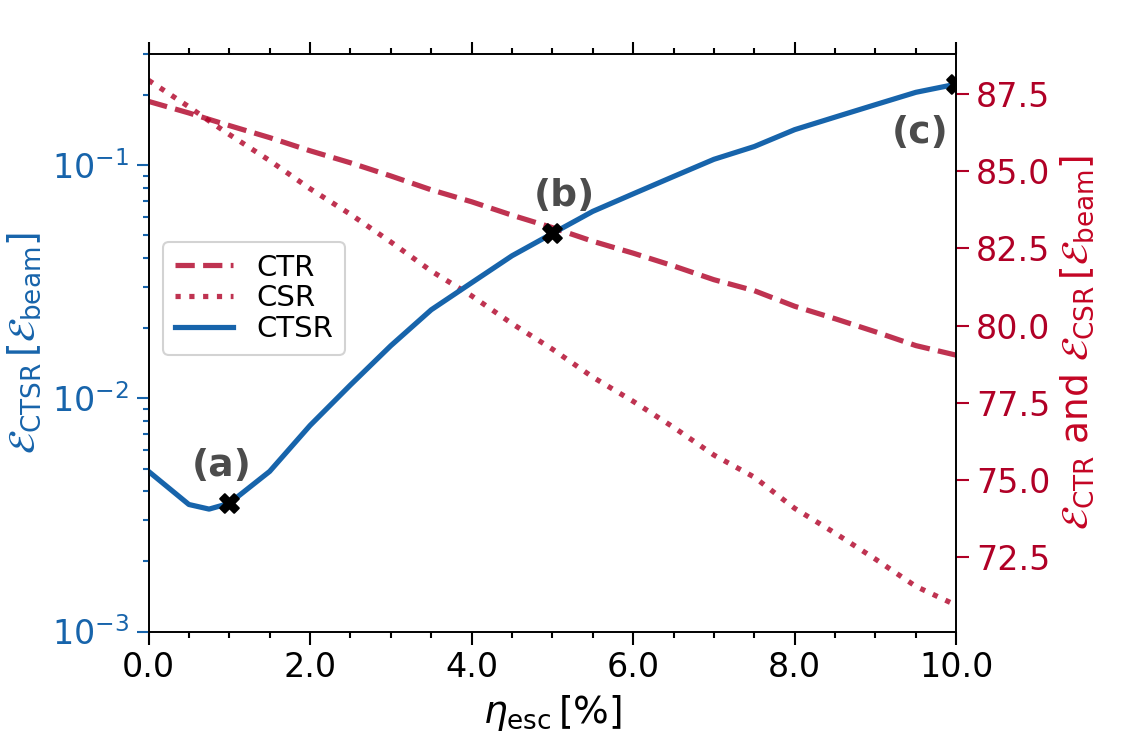}    
    \caption{Radiated THz energy (integrated over solid angles and $0.1-40\,\rm THz$ frequencies) as a function of the fraction of escaping electrons $\eta_{\rm esc}$. The blue solid curve, in $\rm log_{10}$ scale, represents the full radiated energy while the red dashed and dotted curves plot the energy yields of CTR and CSR, respectively. The laser-plasma parameters are those used in Fig.~\ref{fig:ctr:electron_beam:radiat_mech_comp} ($a_L=15$). The radiated spectra of cases (a)-(c) are plotted in Fig.~\ref{fig:ctr:escaping_electrons:spectra}.
    }
    \label{fig:ctr:escaping_electrons:yield}
\end{figure}

As seen previously, very efficient compensation of the radiated THz fields takes place when the fast electrons are rapidly pulled back into the target, as happens under the considered interaction conditions. We have so far assumed that all of the fast electrons are drawn back into the target, yet it is well known that a higher-energy fraction of them can escape the sheath potential \citep{Link:pop:2011, Rusby:hplse_2019}. If these escaping electrons are numerous enough, their uncompensated transition radiation may dominate the total radiation yield. Here, we will assess the impact on the overall THz radiation of an increasing fraction ($\eta_{\rm esc}$, considered as a free parameter) of escaping electrons.

\begin{figure*}
 \centering
 \includegraphics[width=0.99\linewidth]{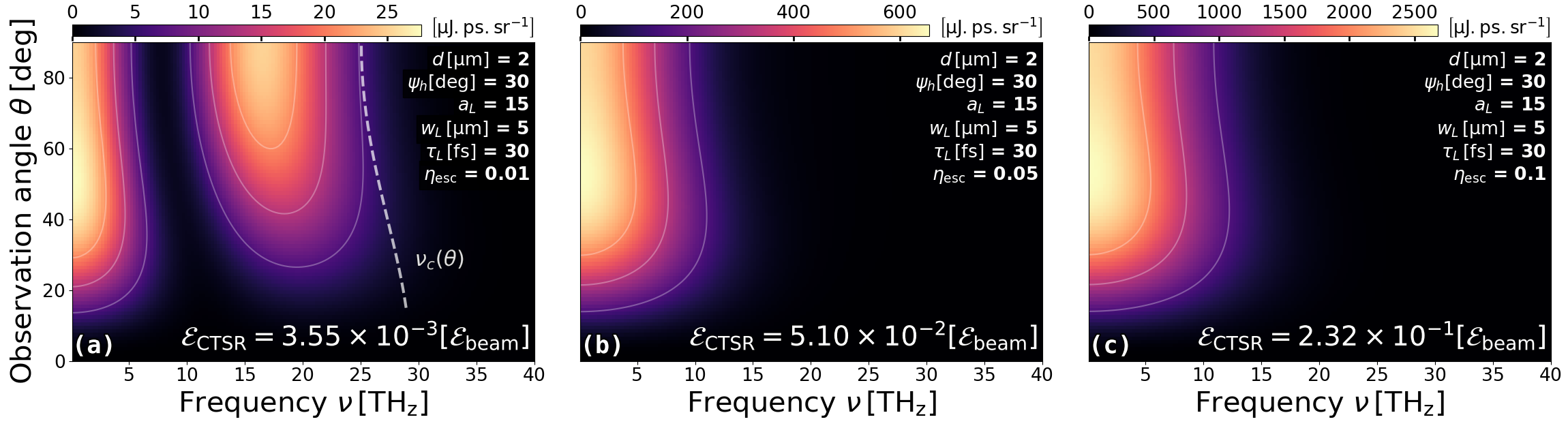}
 \caption{Frequency-angle spectra for various fractions of escaping electrons: (a) $\eta_{\rm esc} = 1\,\%$, (b) $\eta_{\rm esc} = 5\,\%$ and (c) $\eta_{\rm esc} = 10\,\%$, corresponding to cases (a)-(c) in Fig.~\ref{fig:ctr:escaping_electrons:yield}. The other laser-plasma parameters are those used in Fig.~\ref{fig:ctr:electron_beam:radiat_mech_comp} ($a_L=15$).
 The cutoff frequency $\nu_c(\theta)$ is obtained from Eqs.~\eqref{equ:ctr:cutoff_frequency} and \eqref{equ:ctr:electron_beam:Contrast} with $\eta_{\rm coh} = 3 \times 10^{-3}$.}
 \label{fig:ctr:escaping_electrons:spectra}
\end{figure*}

To this purpose, we introduce the cutoff energy $u_c(\eta_{\rm esc}) \equiv p_c(\eta_{\rm esc})/m_e c$ beyond which the electrons escape the target, which fulfills
\begin{align}
    \eta_{\rm esc} &= \int_{u_{\rm c}}^{\infty} g_u(u) u^2 \dd u \nonumber \\
    &= e^{-u_{\rm c}/\Delta u} (2 \Delta u^2 + u_{\rm c}^2 + 2 u_{\rm c} \Delta u )/2 \Delta u^2 \,,
\end{align}
where $\eta_{\rm esc}$ is usually estimated to be in the percent range \citep{Rusby:hplse_2019}.
To simplify, we assume that the escaping electrons (characterized by $u \ge u_{\rm c}$) keep on propagating ballistically after exiting the target backside, and hence only emit a single burst of CTR [stage (iii) in Fig.~\ref{fig:ctr:illustration:target_thin_foil}]. Equations~\eqref{equ:ctr:electron_beam:radiated_energy_Fourier_integrated}, \eqref{equ:ctr:electron_beam:radiated_energy_CTR} and \eqref{equ:ctr:electron_beam:radiated_energy_CSR} can thus be recast as ($\nubar \equiv \nu m_e c / e E_0$):
\begin{widetext}
\begin{align}
    \label{equ:ctr:electron_beam:radiated_energy_CTSR_pc}
    \frac{\partial^2 \SpectralEnergy_{\rm CTSR}}{\partial \nu \partial \Omega}(\rhat, \nu) & = \frac{N_h^2q^2}{2 \pi^2 \eps_0 c } \left\vert \int_0^{u_c} \psiphiIntegral  g(\u) F(\theta, \nu,\u) i \pi \nubar \Bigl[ \widebar{\A}^\nameRealParticle (\rhat,\nu) - \widebar{\A}^\nameImageParticle (\rhat,\nu) \Bigr] \dd^3 \u \right. \nonumber\\
    &\hspace{28pt} + \frac{1}{2} \left. \int_{u_c}^{\infty} \psiphiIntegral g(\u) F(\theta, \nu,\u) \left[ \frac{\rhat \times \fatbeta_{\nameRear}^{-}}{1 - \fatbeta_{\nameRear}^{-} \cdot \rhat} - \frac{\rhat \times \fatbeta_{\nameRear}^{+}}{1 - 
    \fatbeta_{\nameRear}^{+} \cdot \rhat} \right] e^{-i \Theta_{\nameRear}(\u)} \, \dd^3 \u \right\vert^2 \,,\\[20pt]
    \label{equ:ctr:electron_beam:radiated_energy_CTR_pc}
    \frac{\partial^2 \SpectralEnergy_{\rm CTR}}{\partial \nu \partial \Omega}(\rhat, \nu) & = \frac{N_h^2q^2}{2 \pi^2 \eps_0 c } \left\vert \int_0^{u_c} \psiphiIntegral g(\u) F(\theta, \nu,\u) \AInterfaces \, \dd^3 \u \right. \nonumber\\
    &\hspace{28pt} + \frac{1}{2} \left. \int_{u_c}^{\infty} \psiphiIntegral g(\u) F(\theta, \nu,\u) \left[ \frac{\rhat \times \fatbeta_{\nameRear}^{-}}{1 - \fatbeta_{\nameRear}^{-} \cdot \rhat} - \frac{\rhat \times \fatbeta_{\nameRear}^{+}}{1 - 
    \fatbeta_{\nameRear}^{+} \cdot \rhat} \right] e^{-i \Theta_{\nameRear}(\u)} \, \dd^3 \u \right\vert^2 \,,\\[20pt]
    \label{equ:ctr:electron_beam:radiated_energy_CSR_pc}
    \frac{\partial^2 \SpectralEnergy_{\rm CSR}}{\partial \nu \partial \Omega}(\rhat, \nu) & = \frac{N_h^2q^2}{2 \pi^2 \eps_0 c} \left\vert \int_0^{u_c} \psiphiIntegral g(\u) F(\theta, \nu,\u) \Biggl\{ \vphantom{\int} i \pi \nubar \Bigl[ \widebar{\A}^\nameRealParticle (\rhat,\nu) - \widebar{\A}^\nameImageParticle (\rhat,\nu) \Bigr] - \AInterfaces \Biggr\} \,\dd^3 \u \right\vert^2 \,.
\end{align}
\end{widetext}

Figure~\ref{fig:ctr:escaping_electrons:yield} shows the evolution of the radiation yield in the $0.1-40\,\rm THz$ frequency range when $\eta_{\rm esc}$ increases from $0\,\%$ to $10\,\%$. We observe that the CTR yield progressively diminishes with $\eta_{\rm esc}$ while the CSR yield decreases twice as fast.

The decreasing CTR yield is due to the fact that, as $\eta_{\rm esc}$ rises, fewer and fewer electrons emit CTR at time $t= t_{\nameTraj}$ while the number of electrons emitting CTR at $t= t_{\nameRear}$ remains unchanged. Similarly, the CSR yield drops because fewer particles are decelerated. Therefore, when increasing $\eta_{\rm esc}$, one source of CTR is preserved [stage (iii) in Fig.~\ref{fig:ctr:illustration:target_thin_foil}] whereas the whole source of CSR is degraded [stage (iv) of Fig.~\ref{fig:ctr:illustration:target_thin_foil}], which explains the trends seen in Fig.~\ref{fig:ctr:escaping_electrons:yield}.

In parallel, the total (CTSR) THz yield weakly varies (reaching a slight minimum around $\eta_{\rm esc} \lesssim 1\,\%$) for $\eta_{\rm esc} \lesssim 2\,\%$ but rapidly rises beyond this value, by $\sim 65\times$ between $\eta_{\rm esc} = 1\,\%$ and $\eta_{\rm esc} = 10\,\%$. Note that our neglect of radiative losses becomes questionable at $\eta_{\rm esc} = 10\,\%$, for which the yield attains $\Energy_{\rm CTSR} \simeq 0.2\,\Energy_{\rm beam}$.

Figure~\ref{fig:ctr:escaping_electrons:spectra} displays the frequency-angle spectra associated with $\eta_{\rm esc}=1\,\%$, $5\,\%$ and $10\,\%$, corresponding, respectively, to markers (a), (b) and (c) in Fig.~\ref{fig:ctr:escaping_electrons:yield}. When $\eta_{\rm esc}$ rises from zero, the spectrum develops an increasingly prominent low-frequency structure, with a large-angle tail around $\theta \simeq 40-70\deg$, quite above the electron beam's divergence ($\Psi_h=30\deg$). For thin ($d=2\,\rm \mu m$) targets, the hierarchy between CTSR from the confined electrons and CTR from the escaping electrons is reversed at around $\eta_{\rm esc} \simeq 1\,\%$ [Fig.~\ref{fig:ctr:escaping_electrons:spectra}(a)], leading to a doubly peaked spectrum, yet with an integrated energy ($\simeq 3.5 \times 10^{-3}\, \Energy_{\rm beam} \simeq 2.0\,\rm mJ$) comparable with that obtained for $\eta_{\rm esc}=0\%$ (Fig.~\ref{fig:ctr:electron_beam:radiat_mech_comp}). When $\eta_{\rm esc} > 1\%$ [Fig.~\ref{fig:ctr:escaping_electrons:spectra}(b,c)], the low-frequency signal from the escaping electrons largely prevails.

\begin{figure*}
    \includegraphics[width=0.99\linewidth]{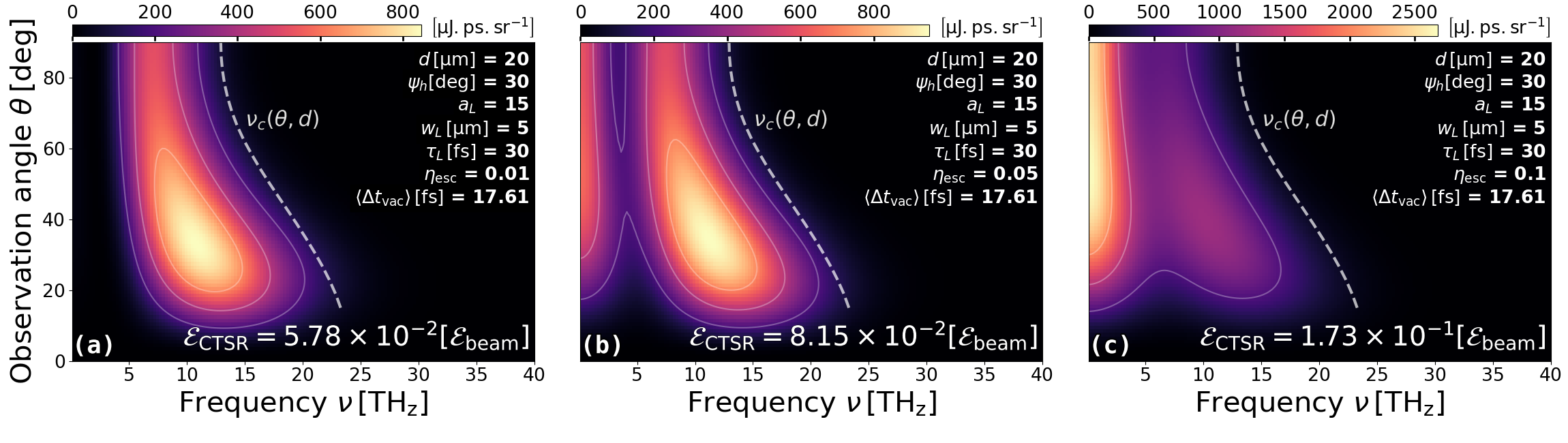}
    \caption{Frequency-angle spectra for a $20\,\rm \mu m$ thick foil and various fractions of escaping electrons: (a) $\eta_{\rm esc} = 1\,\%$, (b) $\eta_{\rm esc} = 5\,\%$ and (c) $\eta_{\rm esc} = 10\,\%$. The other laser-plasma parameters are those used in Fig.~\ref{fig:ctr:electron_beam:radiat_mech_comp} ($\Energy_L=2.8\,\rm J$, $\Energy_{\rm beam}=0.56\,\rm J$).
    The cutoff frequency $\nu_c(\theta,d)$, plotted as a white dashed line, is obtained from Eqs.~\eqref{equ:ctr:electron_beam:definition_attenuation_factor} and \eqref{equ:ctr:electron_beam:Contrast} with $\eta_{\rm coh} = 3 \times 10^{-3}$.}
    \label{fig:ctr:target_thickness:spectra}
\end{figure*}

\subsection{Influence of the target thickness}
\label{subsec:ctr:target thickness}

As the electron beam propagates ballistically through the target, it spreads transversely and longitudinally depending on its momentum distribution function. The increased electron dilution at the backside of a thicker target also entails a weaker sheath field, thus allowing the electrons to propagate in vacuum over larger distances and longer times. As seen in Sec.~\ref{subsec:ctr:return_time}, this tends to reduce the respective yields and frequency ranges of CTR and CSR but also, and above all, to hamper the field cancellation of these two emissions, hence causing a net increase in the CTSR yield. 

To illustrate this behavior, we plot in Fig.~\ref{fig:ctr:target_thickness:spectra} the total THz spectra obtained for $d=20\,\rm \mu m$ and an escaping electron fraction ranging from $1\,\%$ to $10\,\%$. The main observations are that the radiated energy is $\sim 16\times$ higher than at $d=2\,\rm \mu m$ and that the spectrum remains essentially unchanged up to $\eta_{\rm esc} \simeq 2-3\,\%$. This means that the THz radiation is less sensitive to the (unknown) fraction of escaping electrons in thicker foils. When $\eta_{\rm esc} = 10\,\%$, the signal from the recirculating electrons is admittedly less intense than the (lower-frequency) one due to the escaping electrons, but still visible. Moreover, compared to the case of $d=2\,\rm \mu m$ and $\eta_{\rm esc} = 0\,\%$, the spectra associated with $d=20\,\rm \mu m$ have shifted to lower angles ($\sim 25-45\deg$) and frequencies ($\sim 10\,\rm THz$). This behavior results from the lower coherence of the electron beam  (characterized by $C_{\rm coh}$, see Sec.~\ref{subsec:ctr:interplay}) at the rear side of the target. To illustrate the effect of coherence loss, we overlay in Fig.~\ref{fig:ctr:target_thickness:spectra} the cutoff frequency $\nu_c(\theta,d)$ (white dashed line) as obtained by numerically solving Eqs.~\eqref{equ:ctr:electron_beam:definition_attenuation_factor} and \eqref{equ:ctr:electron_beam:Contrast} for $\eta_{\rm coh} = 3\times 10^{-3}$. This cutoff frequency captures fairly well the high-$\nu$ shape of the CTSR spectrum.

The prediction of a THz yield increasing with the target thickness between $d=2\,\rm \mu m$ and $d=20\,\rm \mu m$ and for $\eta_{\rm esc}=1-5\,\%$ [Figs.~\ref{fig:ctr:escaping_electrons:spectra}(a,b) and ~\ref{fig:ctr:target_thickness:spectra}(a,b)] may seem dubious given the absence of supporting experimental evidence, but should be considered along with the $\eta_{\rm esc}=10\,\%$ case. The latter indeed suggests that, for a given fraction of escaping electrons $\eta_{\rm esc}$, the THz yield will eventually drop with the target thickness as CTSR, which is increasingly dominated by CTR when $\eta_{\rm esc}>1\,\%$, weakens due to degraded beam coherence.

\subsection{Influence of the laser focusing}
\label{subsec:ctr:laser_field}
 
We now examine how the THz radiation depends on the laser pulse intensity ($\propto\, a_L^2$) at fixed laser energy ($\propto\, a_L^2 w_L^2$). To do so, we vary the laser field strength $a_L$ while adjusting the laser spot size $w_L\,\propto\, 1/a_L$ and keeping the laser-to-hot-electron conversion efficiency fixed to $\eta_h=0.2$. Since $\langle \gamma_h \rangle\,\propto\, a_L$, lowering $a_L$ causes the number of hot electrons $N_h\,\propto\, \Energy_L/\langle \gamma_h \rangle$ to rise, leading to stronger separate CTR and CSR yields. Predicting the net effect on CTSR, though, is not straightforward as it is extremely sensitive to the normalized return time $\langle \widebar{\Delta t_r} \rangle \,\propto\, \Delta u \,\propto\, a_L$, as shown by Eqs.~\eqref{equ:ctr:electron_beam:radiated_energy_Fourier_integrated} - \eqref{equ:ctr:electron_beam:definition_of_A_rel} and Secs.~\ref{subsec:ctr:interplay} and \ref{subsec:ctr:return_time}.

\begin{figure*}
    \centering
    \includegraphics[width=0.99\linewidth]{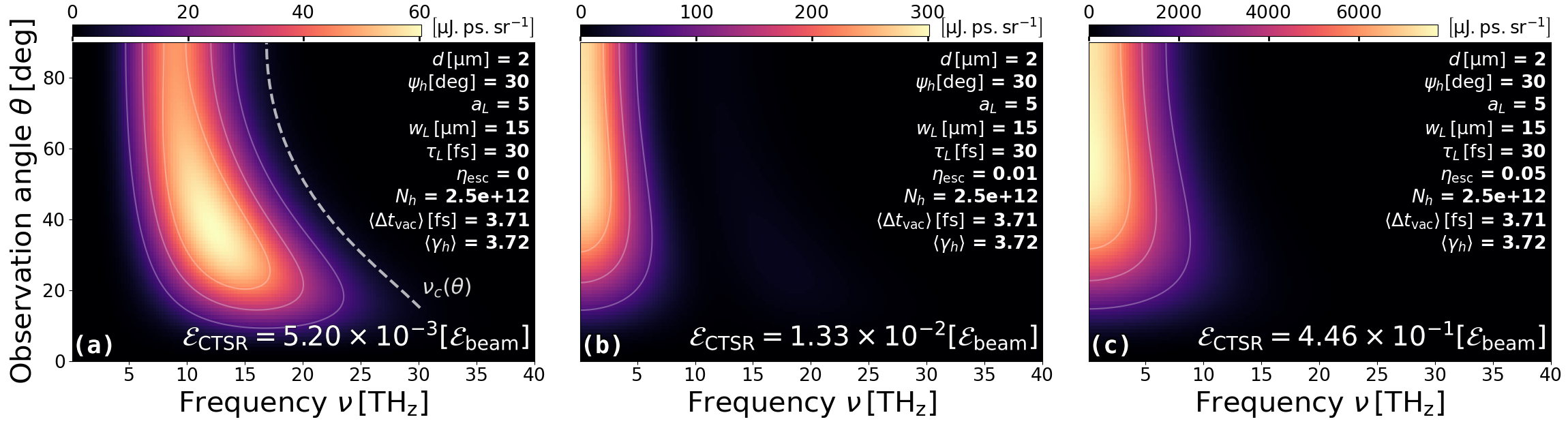}
    \includegraphics[width=0.99\linewidth]{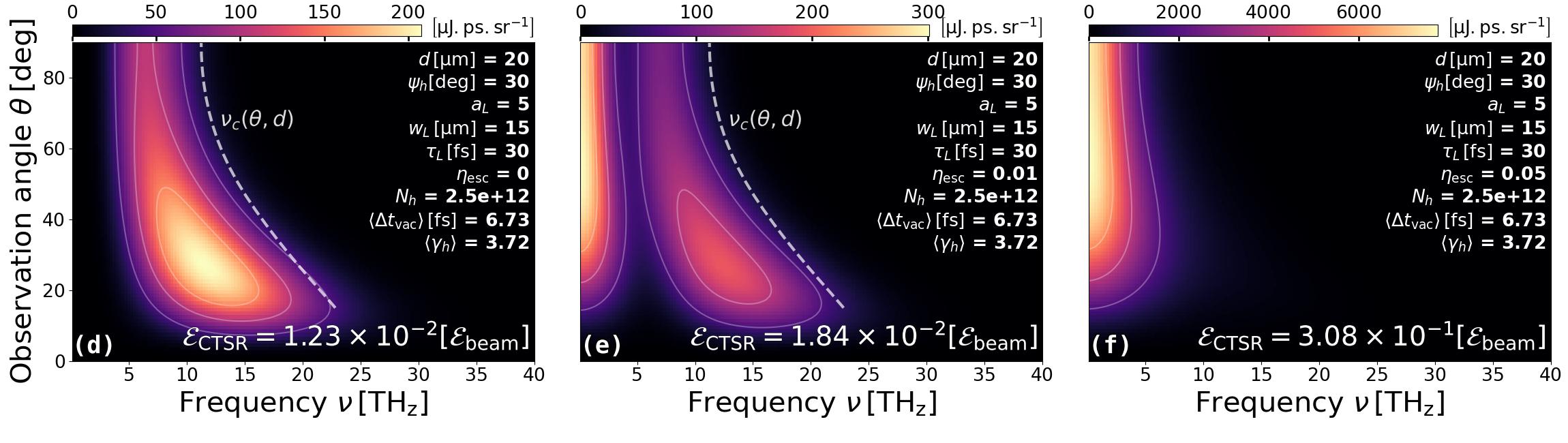}
    \caption{Frequency-angle spectra (in $\rm \mu J\,ps\,sr^{-1}$ units) for (top) $d=2\,\rm\mu m$ and (bottom) $d=20\,\rm\mu m$ at $a_L=5$ and fixed laser pulse energy $\Energy_L=2.8\,\rm J$ ($\Energy_{\rm beam} = 0.56\,\rm J$).
    The fraction of escaping electrons varies from left to right: (a,d) $\eta_{\rm esc} = 0\,\%$, (b,e) $\eta_{\rm esc} = 1\,\%$ and (c,f) $\eta_{\rm esc} = 5\,\%$.
    Apart from $d$, $a_L$ and $w_L \propto a_L^{-1}$, the other parameters are as in Fig. \ref{fig:ctr:electron_beam:radiat_mech_comp}. 
    Panels (b,c) (resp. (e,f)) are to be compared directly to their $a_L=15$ equivalent [Figs.~\ref{fig:ctr:escaping_electrons:spectra}(a,b) and \ref{fig:ctr:target_thickness:spectra}(a,b)].
    The cutoff frequencies (a) $\nu_c(\theta)$ and (d,e) $\nu_c(\theta,d)$ are computed from, respectively, Eq.~\eqref{equ:ctr:cutoff_frequency} and Eqs.~\eqref{equ:ctr:electron_beam:definition_attenuation_factor}-\eqref{equ:ctr:electron_beam:Contrast} with $\eta_{\rm coh} = 1 \times 10^{-3}$.
    }
    \label{fig:ctr:aL:spectra}
\end{figure*}

Figure~\ref{fig:ctr:aL:spectra} plots the THz radiated spectra for (top) $d=2\,\rm\mu m$ and (bottom) $d=20\,\rm\mu m$, the laser pulse energy being set to $\Energy_L=2.8\,\rm J$, the laser amplitude to $a_L=5$, and thus the waist to $w_L=15\,\rm\mu m$. Panels (b,c) [(e,f)] are to be compared directly to their $a_L=15, w_L = 5\,\rm\mu m$ [Fig.~\ref{fig:ctr:escaping_electrons:spectra}(a,b) [resp. Fig.~\ref{fig:ctr:target_thickness:spectra}(a,b)] equivalent. For $d=2\,\rm\mu m$, and assuming full refluxing ($\eta_{\rm esc}=0\,\%$), the total radiated energy  ($\Energy_{\rm CTSR} \simeq 5.2\times 10^{-3}\,\Energy_{\rm beam}$) is almost identical to that achieved at $a_L=15$, yet the radiation is more sensitive to the escaping electrons: their contribution becomes overwhelmingly dominant even when their fraction is as low as $1\,\%$, in which case the radiation yield is more than doubled ($\Energy_{\rm CTSR} \simeq 1.3\times 10^{-2}\,\rm \Energy_{\rm beam}$). 
Again, when $\eta_{\rm esc}$ is sufficiently low, increasing the target thickness from $d=2\,\rm \mu m$ to $d=20\,\rm\mu m$ significantly enhances the radiated energy, albeit to a lesser degree than at $a_L=15$. The contribution of the recirculating electrons is then still significant at $\eta_{\rm esc} = 1\,\%$, but no longer so at $\eta_{\rm esc} = 5\,\%$. Note, however, that for both $d=2\,\rm \mu m$ and $d=20\,\rm \mu m$, the model's predictions become questionable when $\eta_{\rm esc} \gtrsim 5\,\%$ since $\Energy_{\rm CTSR}$ then exceeds $\sim 0.3\,\Energy_{\rm beam}$.

Interestingly, the $\sim 10^{-2}$ beam-to-THz conversion efficiency predicted for $a_L=5$ and $\eta_{\rm esc} = 1\,\%$ appears to be roughly consistent with the $\sim 1\,\rm mJ$ yield reported in Ref.~\onlinecite{Herzer:njp:2018} under comparable interaction conditions ($a_L \simeq 5$, $\tau_L \simeq 30\,\rm fs$, $w_L \simeq 10\,\rm \mu m$), but with a thicker foil ($d=5\,\rm \mu m$) and a lower cutoff frequency ($\nu < 9\,\rm THz$).  

Overall, our model pinpoints the critical influence of some of its parameters, notably the sheath field strength and the fraction of escaping electrons. Yet, to our knowledge, most experiments to date have characterized THz emissions from laser-foil interactions over restricted frequency and angular ranges\cite{Jin:pre:2016, Herzer:njp:2018}, with the notable exception of Ref.~\onlinecite{Gopal:pre:2019} where the full THz distribution was diagnosed, but with an obliquely incident laser pulse and no variation in the target thickness was reported.



\section{Radiation from the expanding plasma}
\label{sec:sr:radiation_expanding_plasma}

We now address the THz radiation that is subsequently emitted by the electron-proton plasma expanding due to the fast-electron-induced sheath field. Compared to previous efforts \cite{Kuratov:qe:2016, Herzer:njp:2018, Woldegeorgis:pre:2019, Liao:pnas:2019, Liao:prx:2020}, our modeling will hinge on a more realistic description of the plasma acceleration, taking account of the time-decreasing areal charge at the accelerated ion front and of the adiabatic electron cooling taking place in thin foils.

\subsection{General formalism}
\label{subsec:sr:formalism}

A convenient formula to describe the far-field plasma expansion radiation (PER) is~\cite{McDonald:ajp:1997, Panofsky:cc:2005}
\begin{multline}
    \label{equ:sr:radiated_energy}
    \frac{\partial^2 \SpectralEnergy_{\rm PER}}{\partial \nu \partial \Omega}(\rhat, \nu) = \frac{1}{8\pi^2 \varepsilon_0 c^3} \\
    \times \left \vert \iint \rhat \times \retardedTime{\frac{ \dpa \j (\r', t) }{\dpa t} } e^{-2 i \pi \nu t}  \dd t \dd^3 r' \right \vert^2 \,,
\end{multline}
where $\j$ is the (electron-proton) plasma current density which will be estimated below. We will assume, for simplicity, that the plasma is accelerated along the target normal only. Thus, $\j(\r,t) = j(\r,t) \zhat$ and the radiated spectrum associated with Eq.~\eqref{equ:sr:radiated_energy} reads
\begin{multline}
    \label{equ:sr:radiated_energy_djdt}
    \frac{\partial^2 \SpectralEnergy_{\rm PER}}{\partial \nu \partial \Omega}(\theta, \nu) = \frac{\sin^2 \theta}{8\pi^2 \eps_0 c^3} \\
    \times \left\vert \iint \retardedTime{\frac{\dpa j}{\dpa t}(\r',t)} e^{-2 i \pi \nu t} \,\dd t \,\dd^3 \r' \right \vert^2  \,.
\end{multline}
Performing the change of variable $t \to \tret$ leads to 
\begin{multline}
    \label{equ:sr:radiated_energy_djdt:retarded}
    \frac{\partial^2 \SpectralEnergy_{\rm PER}}{\partial \nu \partial \Omega} (\theta,\nu) = \frac{\sin^2 \theta}{8\pi^2 \eps_0 c^3} \\
    \times \left| \iint \frac{\dpa j}{\dpa t} (\r',t') e^{-2 i \pi \nu \left( t' - \frac{\rhat \cdot \r'}{c} \right)} \, \dd t' \,\dd^3 \r' \right|^2 \,.
\end{multline}
This formula is the basis of the PER model constructed below.

\subsection{Plasma expansion in the isothermal and adiabatic regimes}
\label{subsec:sr:tnsa_regimes}

To evaluate the current density $j(\r,t)$, we need to describe the dynamics of the expanding plasma. To do so, we start by considering the one-dimensional (1D) model proposed in Ref.~\onlinecite{Mora:prl:2003}, which applies to a collisionless plasma accelerated towards vacuum by the sheath field created by an isothermal electron population. According to this model, the plasma ion profile, initialized as a step function, retains a sharp front located at $z_{\rm f,iso}(t)$ and moving at the velocity
\begin{equation}
    \label{equ:sr:fastest_protons:velocity}
    \dot{z}_{\rm f,iso}(t) \equiv \beta_{\rm f,iso}(t) c \simeq 2 c_{\rm s0} \ln \left(\tau + \sqrt{\tau^2 + 1} \right)\,,
\end{equation}
where $c_{\rm s0} \simeq \sqrt{\langle \Energy_h \rangle/m_p}$ is the ion acoustic velocity, $m_p$ the proton mass, $\tau = \omega_{pi} t /\sqrt{2 e_N}$ the normalized time, $\omega_{pi} = \sqrt{n_{\rm hr0}e^2/m_p \eps_0}$ the ion plasma frequency, $e_N \equiv \exp(1)$, and $n_{\rm hr0}$ the initial electron density at the target rear side.
We estimate $n_{\rm hr0}$ as in Eq.~\eqref{equ:ctr:nhr0} but consider in addition a folding term $(1 + c \tau_L / 2d)$ to describe the electron accumulation in thin foils during the laser irradiation~\cite{McKinnon:prl:2002}:
\begin{equation}
    \label{equ:sr:electronic_density:rear_definition}
    n_{\rm hr0} = \frac{\eta_h I_L}{m_e c^3 \langle (\gamma_h-1) \beta_z \rangle} \left(\frac{w_h(0)}{w_h(d)}\right)^2 \left( 1 + \frac{c\tau_L }{2d}\right)\,.
\end{equation}
When the longitudinal extent $\sim c\tau_L $ of the fast electron bunch is short compared to the target thickness, i.e., when $c \tau_L \ll 2d$, the folding term approaches unity and the electron density is that given by Eq.~\eqref{equ:ctr:nhr0}. Conversely, when $c \tau_L \gtrsim 2d$, the fast electrons recirculate several times before the pulse ends and the folding term increases the electron density accordingly.

The approximation of isothermal hot electrons ceases to be valid when the rarefaction waves coming from the two sides of the foil have converged to its center, which occurs at a time $t_{\rm ad} \simeq d/2 c_{\rm s0}$~\cite{Mora:pre:2005}.  The adiabatic cooling experienced by the fast electrons at $t> t_{\rm ad}$ precipitates the decay of the sheath field, so that the ions eventually reach a maximum velocity. In a 1D geometry, this maximum velocity is predicted to be~\cite{Mora:pre:2005} 
\begin{equation}
    \label{equ:definition_v_adiabatic}
    \dot{z}_{\rm f,ad} \simeq 2 c_{\rm s0} \ln \left( 0.32 d/\lambda_{D0} + 4.2 \right) \,,
\end{equation}
where 
\begin{equation}
    \label{equ:definition_lambda_d0}
    \lambda_{D0} = c_{\rm s0}/\omega_{pi} = \sqrt{\eps_0 \langle \Energy_h \rangle / n_{\rm hr0} e^2 }
\end{equation} is the initial Debye length of the hot electrons.

To model the dynamic transition between the isothermal and adiabatic (cooling) expansion regimes, we use the simple interpolation formula proposed in Ref.~\onlinecite{Ferri:pop:2018}:
\begin{equation}
    \label{equ:definition_vf1D}
    \dot{z}_{\rm f,1D}(t) \simeq \left[ \dot{z}^{-2}_{\rm f,iso}(t) + \dot{z}_{\rm f,ad}^{-2} \right]^{-1/2}.
\end{equation}
As the time-dependent sheath field (as seen by the front ions) fulfills $\ddot{z}_{\rm f,1D}(t) = eE_{\rm f,\,1D}(t)/m_p$, we deduce
\begin{equation}
    \label{equ:definition_EF1D}
    E_{\rm f,1D}(t) = \frac{m_p}{e} \frac{\ddot{z}_{\rm f,iso}(t)}{{\left[ 1 + \dot{z}^2_{\rm f,iso}(t) /\dot{z}^2_{\rm f,ad} \right]^{3/2}}}\,.
\end{equation}

Next, in order to consider the effect of the transverse spreading of the hot electrons, we note that the electric field initially scales as $\sqrt{n_{\rm hr0}}$ in the purely 1D case~\cite{Mora:prl:2003}, and so apply for $t > \tau_L$ a correction factor $\sqrt{n_{\rm hr}(t)/n_{\rm hr0}} = w_h(d)/w_h(d+c(t-\tau_L))$ to $E_{\rm f,1D}$ above. Finally, following Refs.~\onlinecite{Brantov:prstab:2015, Ferri:pop:2018}, we take into account the expected weakening of the sheath field when the ion front has moved a distance $z_{\rm f}$ greater than its transverse extent $w_h(d)$ [$w_h(z)$ being the transverse electron beam size given by Eq.~\eqref{equ:ctr:whz}]. The sheath field is then expected to decay in time as $z^{-2}_{\rm f}(t)$ in a realistic 3D geometry~\cite{Brantov:prstab:2015, Ferri:pop:2018}. 

All these considerations motivate the following approximate expression of the sheath field
\begin{align}
\label{equ:definition_Ex3D}
    E_{\rm f,3D}(t) &=
    \begin{cases}
        E_{\rm f,1D}(t) \left[1\! +\! \frac{z^2_{\rm f,3D}(t)}{w_h^2(d)} \right]^{-1} & \hspace{-5pt} , t < \tau_L \\[6pt]
        \frac{E_{\rm f,1D}(t) w_h(d)}{w_h(d + c(t-\tau_L))} \left[1\! +\! \frac{z^2_{\rm f,3D}(t)}{w_h^2(d)} \right]^{-1} & \hspace{-5pt} ,t \geq \tau_L 
    \end{cases}\\
    \label{equ:definition_Ex3D_bis}
    \ddot{z}_{\rm f,3D}(t) &= \frac{e}{m_p} E_{\rm f,3D}(t)\,.
\end{align}
The above equations are solved numerically to obtain the time-varying position, velocity and acceleration of the (fastest) front ions.

\begin{figure}
    \centering
    \includegraphics[width=0.9\linewidth]{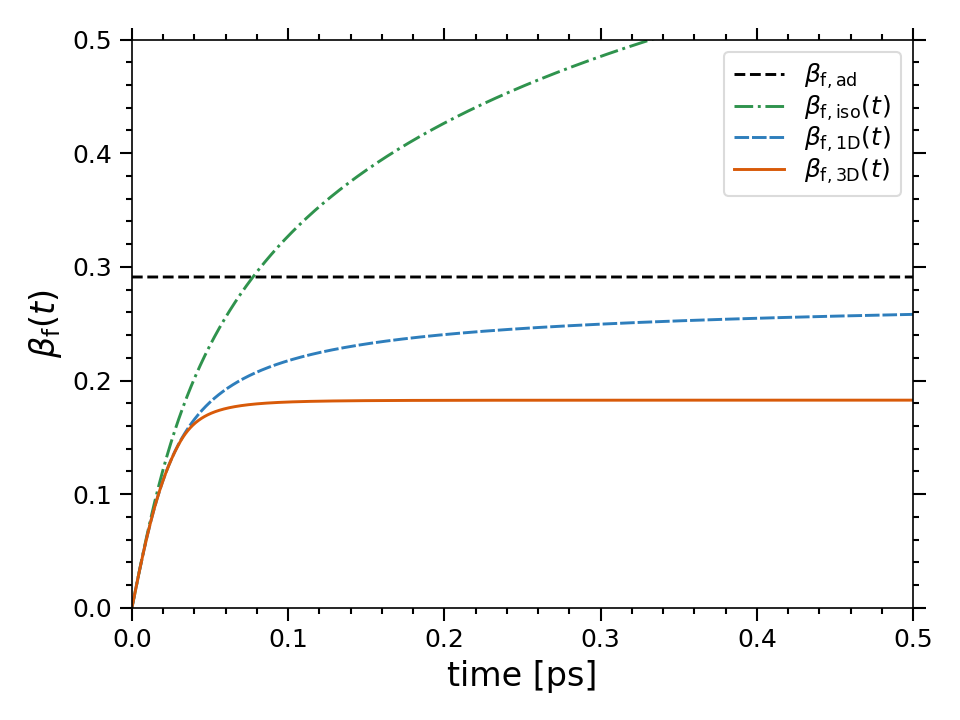}
    \caption{Time evolution of the ion front velocity as predicted by various models. Green dotted dashed curve: isothermal model [$\beta_{\rm iso,1D}$ from Eq.~\eqref{equ:sr:fastest_protons:velocity}]. Blue dashed curve: mixed 1D isothermal/adiabatic model [$\beta_{\rm f,1D}$ from Eq.~\eqref{equ:definition_vf1D}]. Orange solid curve: mixed 3D isothermal/adiabatic regime [$\beta_{\rm f,3D}$ from Eq.~\eqref{equ:definition_Ex3D_bis}]. Black dashed curve: maximum velocity in the adiabatic regime [$\beta_{\rm f,ad}$ from Eq.~\eqref{equ:definition_v_adiabatic}]. The laser-plasma parameters are those used in Fig.~\ref{fig:ctr:electron_beam:radiat_mech_comp}.}
    \label{fig:SR:evolution_ferri2018}
\end{figure}

As an example, Fig.~\ref{fig:SR:evolution_ferri2018} compares the normalized proton front velocities, $\beta_{\rm f} =\dot{z}_{\rm f}/c$, as obtained from the above models. The laser-plasma parameters ($a_L=15$, $w_L=5\,\rm \mu m$, $\tau_L = 30\,\rm fs$, $d=2\,\rm \mu m$) are the same as in Fig.~\ref{fig:ctr:electron_beam:radiat_mech_comp}.  
Plotted are the maximum adiabatic velocity \cite{Mora:pre:2005} $\dot{z}_{\rm f, ad}/c$, the unbounded ion velocity in the isothermal~\cite{Mora:prl:2003} regime, $\dot{z}_{\rm f,iso}/c$, and the ion velocity in the mixed isothermal/adiabatic regime without ($\dot{z}_{\rm f,1D}/c$) or with ($\dot{z}_{\rm f,3D}/c$) 3D corrections~\cite{Ferri:pop:2018}. 
The rapid saturation of the ion velocity (reaching a maximum value $\beta_{\rm f,3D} \simeq 0.18$ by $t \simeq 0.1\,\rm ps$) that is predicted by the 3D adiabatic model is particularly manifest.

\subsection{Evaluation of the plasma current density}
\label{subsec:sr:energy_spectrum}

The plasma current density $j(\r,t)$ (resulting from both electron and ion contributions) involved in Eq.~\eqref{equ:sr:radiated_energy_djdt:retarded} is inferred from the charge density \textReferee{$\rho(\r,t) \equiv e \left(Z n_i - n_e \right)  (\r,t)$ ($n_e$ and $n_i$ are the electron and ion number densities)} via the \textsc{1D} charge conservation equation $\dpa_t \rho + \dpa_z j = 0$. As described in Ref.~\onlinecite{Mora:prl:2003}, the density profile of the expanding plasma exhibits two charge-separation regions: one around the mobile ion front ($z\simeq z_{\rm f}$) with negative areal charge density $-\sigma_{\rm f}(t)$ and one around the initial plasma surface ($z\simeq 0$) with positive areal charge density $+\sigma_{\rm f}(t)$. We will neglect the longitudinal extent of these nonneutral layers and model $\rho(\r,t)$ as a sum of two charged planes, centered at $z\simeq z_{\rm f}(t)$ and $z=0$,
\begin{equation}
    \rho(\r,t) = \sigma_{\rm f}(t) \left[\delta(z) - \delta(z-z_{\rm f}(t)) \right] f_{\perp,\textsc{PE}}(\r_\perp) \,.
\end{equation}
Here $f_{\perp,\textsc{PE}}(\r_\perp)$ describes the transverse density profile of the expanding plasma. In principle, this profile should vary in time since the plasma expands both in the longitudinal and transverse directions. For simplicity, we consider a constant transverse distribution of Gaussian shape and same width as that of the recirculating electron beam:
\begin{equation}
    \label{equ:sr:f_perp:definition}
    f_{\perp,\textsc{PE}}(\r_\perp) = e^{-4 \ln 2 (r_\perp /w_h(d))^2} \,.
\end{equation}
In a 1D geometry, the time-varying areal charge density is linked to the sheath field through $\sigma_{\rm f}(t) = \varepsilon_0 E_{\rm f,1D}(t)$. For simplicity, we assume that this relation remains approximately valid in the 3D expansion regime, i.e., 
\begin{equation}
    \sigma_{\rm f}(t) = \varepsilon_0 E_{\rm f,3D}(t) \,,
\end{equation}
with $E_{\rm f,3D}(t)$ as defined by Eq.~\eqref{equ:definition_Ex3D}. This leads to the current density 
\begin{equation}
    \label{equ:sr:compute_j:compute_j_from_rho}
    j(\r,t) = j_\parallel (z,t) f_{\perp,\textsc{PE}}(\r_\perp) \,,
\end{equation}
where
\begin{align}
    \label{equ:sr:compute_j:compute_j_from_rho:part_2}
    j_\parallel (z,t) &= -\int_ {-\infty}^z \frac{\dpa \rho (\r,t)}{\dpa t} \dd z\nonumber \\
    &=- \dot{\sigma}_{\rm f}(t)\left[ \mathcal{H}_0 - \mathcal{H}_{z_{\rm f}(t)} \right] - \sigma_{\rm f}(t) \dot{z}_{\rm f}(t) \delta_{z_{\rm f}(t)} \,.
\end{align}
Here, $\mathcal{H}_{z_{\rm f}(t)} \equiv \mathcal{H}\left( z - z_{\rm f}(t) \right)$ and $\delta_{z_{\rm f}(t)} \equiv \delta \left(z - z_{\rm f}(t) \right)$ denote, respectively, the Heaviside and Dirac delta functions centered on $z = z_{\rm f}(t)$. 
Finally, introducing the transverse spatial Fourier transform
\begin{align}
    \mathcal{F}[f_{\perp,\textsc{PE}}](k_x, k_y) &= \iint_{-\infty}^\infty \dd x \dd y\,f_{\perp,\textsc{PE}}(x, y) \nonumber \\
    &\times e^{-2 i \pi (k_x x + k_y y)} \nonumber  \\
    &= \pi \frac{w_h^2(d)}{4 \ln 2} e^{-\pi^2 \frac{w_h^2(d)}{4 \ln 2} (k_x^2 + k_y^2)} \,,
\end{align}
we can rewrite Eq.~\eqref{equ:sr:radiated_energy_djdt:retarded} as
\begin{align}
    \label{equ:sr:radiated_energy_final}
    &\frac{\partial^2 \SpectralEnergy_{\rm PER}}{\partial \nu \partial \Omega}(\theta, \nu) = \frac{w_h^4(d) \sin^2 \theta}{2 (8 \ln 2)^2\eps_0 c^3} e^{-\left( w_h(d) \sin \theta \frac{\pi \nu}{c} \right)^2/2 \ln 2} \nonumber \\
    &\times \left| \int \mathcal{F} \left[ \frac{\dpa j_\parallel}{\dpa t'} \right] \left( \frac{- \nu \cos \theta}{c} \right)e^{- 2 i \pi \nu t'} \dd t' \right|^2 \,.
\end{align}
where $\mathcal{F} \left[ \frac{\dpa j_\parallel}{\dpa t'} \right]$ is the spatial (along $z$) Fourier transform of the time derivative $\dpa j_\parallel / \dpa t$, evaluated analytically at $k_z = -\nu \cos \theta/c$ as a function of $z_{\rm f}(t)$ and $\sigma_{\rm f}(t)$ (See \appcite{\ref{sec:app:Fourier_transform_of_current_density}}). The time integral in the above expression is then computed numerically.  

In the following sections, we study the variations in the energy yield and spectra of PER with the main laser-plasma parameters.

\subsection{Influence of the laser focusing}
\label{subsec:sr:laser_parameters}

\begin{figure}
    \centering
    \includegraphics[width=0.9\linewidth]{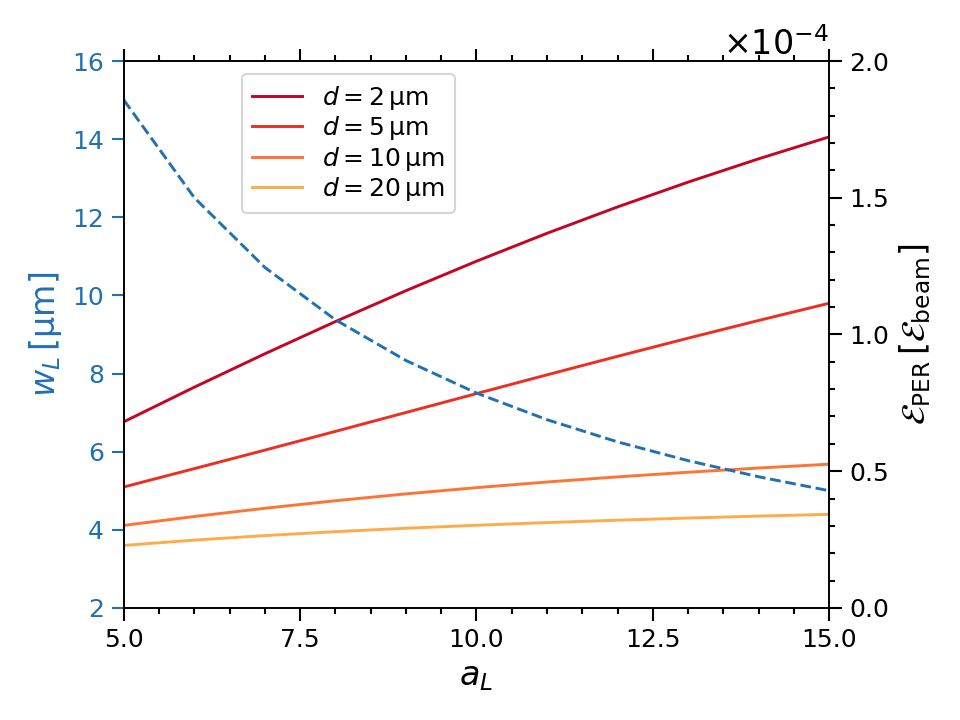}
    \caption{THz energy yield (integrated over $0.1-40\,\rm THz$) of PER as a function of the laser field strength $a_L$ (right axis), for various target thicknesses $d$. The product $a_L w_L$ is kept fixed to ensure a constant laser pulse energy $\Energy_L = 2.8\,\rm J$ (or, equivalently, a constant electron beam energy $\Energy_{\rm beam} = 0.56\,\rm J$). The blue dashed curve plots $w_L$ (left axis). Apart from $a_L$ and $w_L$, the laser-plasma parameters are those used in Fig.~\ref{fig:ctr:electron_beam:radiat_mech_comp}.}
    \label{fig:sr:parametric_studies:aL}
\end{figure}

We first perform a scan over the laser pulse intensity (or, equivalently, degree of laser focusing) at fixed laser pulse energy ($\Energy_L=2.8\,\rm J$) and duration ($\tau_L=30\,\rm fs$), by keeping $a_L w_L$ constant.
Figure~\ref{fig:sr:parametric_studies:aL} plots the evolution of the energy radiated via PER ($\Energy_{\rm PER}$, integrated over $0.1 \le \nu \le 40\,\rm THz$) as a function of $a_L$ (in the range $5\le a_L \le 15$) and for different target thicknesses ($2 \le d \le 20\,\rm \mu m$). As in Sec.~\ref{sec:ctr:model_results}, the laser-to-hot-electron conversion efficiency is set to $\eta_{\rm h} = 0.2$.

First, one observes that the predicted yields lie in the $3\times 10^{-5}- 1.7\times 10^{-4}\,\Energy_{\rm beam}$ range, \emph{i.e.}, at least one order of magnitude below the previous estimates of the energy radiated by the fast electrons alone.  All curves show a monotonic increase with $a_L$. In detail, for $d=2\,\rm \mu m$, $\Energy_{\rm PER}$ is enhanced by a factor of $\sim 2.5$ when $a_L$ is increased from 5 to 15. When $d=5\,\rm \mu m$ ($d=10\,\rm \mu m$), the enhancement factor is of $\sim 2.2$ (resp. $\sim 1.7$).  

\begin{figure}
    \centering
    \includegraphics[width=0.9\linewidth]{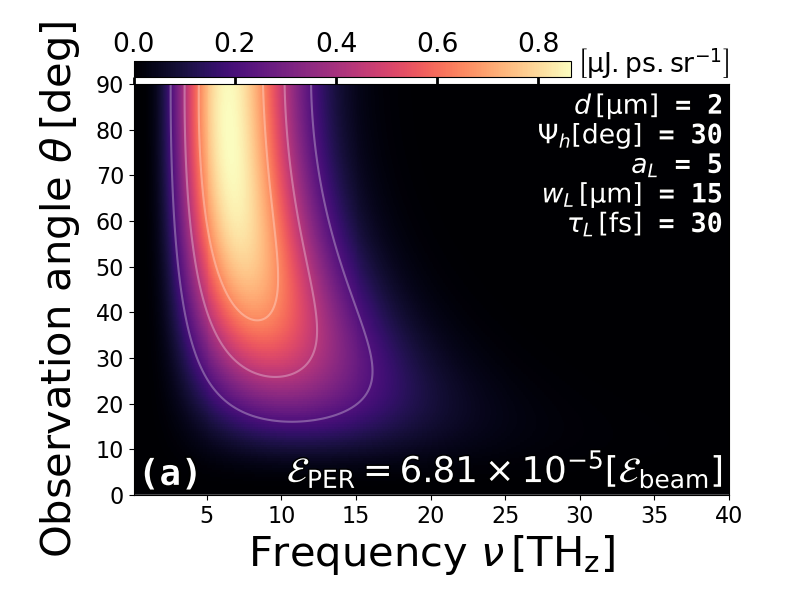}
    \includegraphics[width=0.9\linewidth]{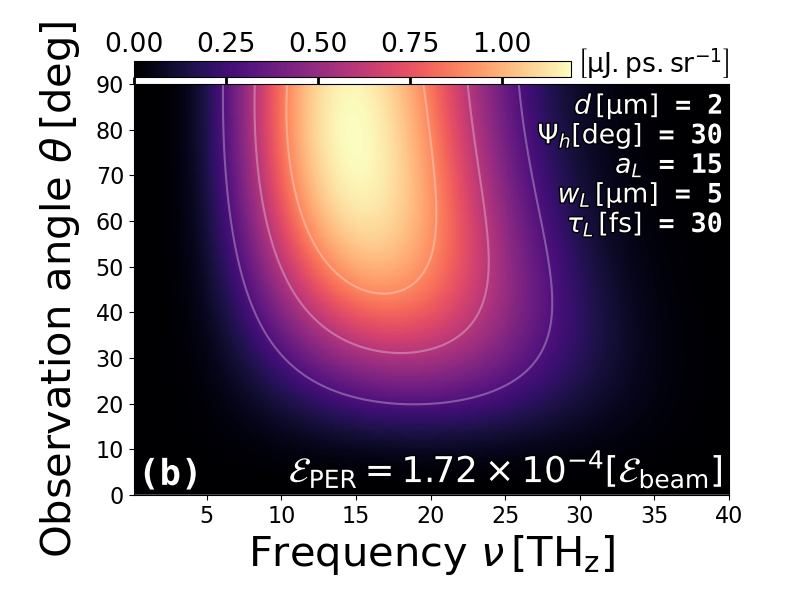}
    \caption{THz energy yield (integrated over $0.1-40\,\rm THz$) of PER for laser field strengths (a) $a_L=5$ and (b) $a_L=15$. The product $a_L w_L$ is kept fixed to ensure a constant laser pulse energy $\Energy_L = 2.8\,\rm J$ (or, equivalently, a constant electron beam energy $\Energy_{\rm beam} = 0.56\,\rm J$). Apart from $a_L$ and $w_L$, the laser-plasma parameters are those used in Fig.~\ref{fig:ctr:electron_beam:radiat_mech_comp}.}
    \label{fig:sr:parametric_studies:aL_5_and_aL_15}
\end{figure}

As displayed in Fig.~\ref{fig:sr:parametric_studies:aL_5_and_aL_15}, intensifying the laser field (from $a_L=5$ to $a_L=15$) by narrowing its spot size (from $w_L=15\,\rm \mu m$ to $5\,\rm \mu m$) broadens and shifts the THz spectrum to higher THz frequencies (from $\sim 5-15\,\rm THz$ to $\sim 10-25\,\rm THz$) because of a faster plasma expansion. This effect is reinforced by the accompanying decrease in transverse width of the sheath field $w_h(d)$ [see Eq.~\eqref{equ:ctr:whz}], which weakens the attenuation factor associated with the Fourier transform of $f_{\perp,\rm PE}$ in Eq.~\eqref{equ:sr:radiated_energy_final}. The two spectra, however, are maximized at large angles $\theta \sim 70-90\deg$, as is typical for nonrelativistic dipole-like radiation. 

\subsection{Influence of the target thickness}
\label{subsec:sr:target_thickness}

\begin{figure}
    \centering
    \includegraphics[width=0.9\linewidth]{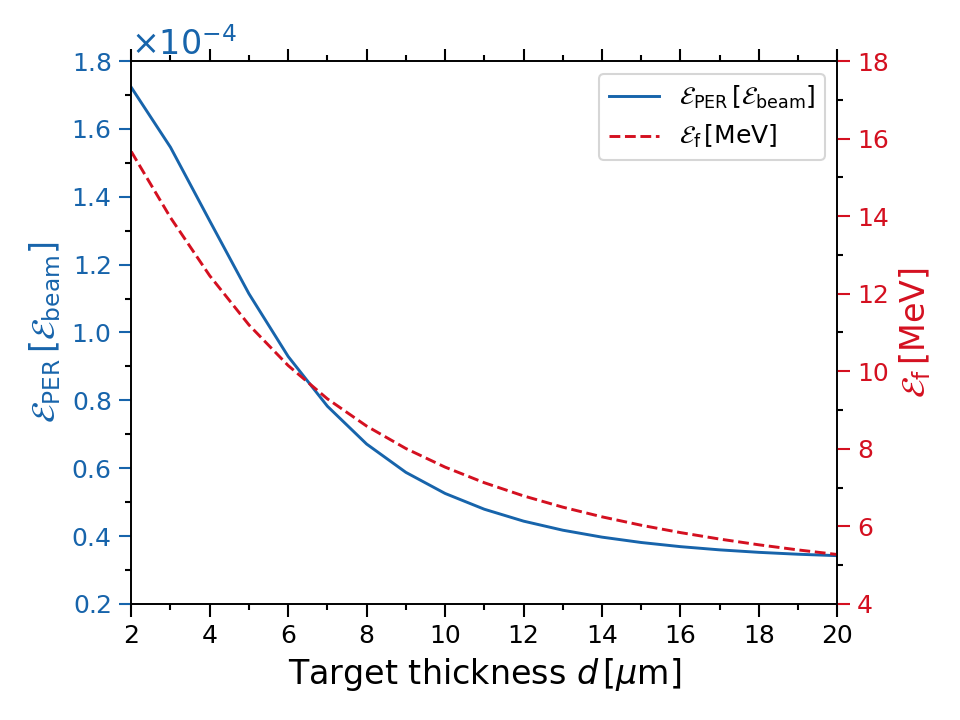}
    \caption{Kinetic energy of the fastest ions (dashed red curve) and energy yield of PER integrated over $0.1-40\,\rm THz$ (solid blue curve) as a function of the target thickness $d \ge 2\,\rm \mu m$. Apart from $d$, the laser-plasma parameters are those used in Fig.~\ref{fig:ctr:electron_beam:radiat_mech_comp} ($a_L=15$).
    }
    \label{fig:sr:parametric_studies:d_1}
\end{figure}

We now vary the target thickness $d$ at fixed laser parameters ($a_L=15$, $w_L=5\,\rm \mu m$, $\tau_L = 30\,\rm fs$, $\Energy_L=2.8\,\rm J$). Figure~\ref{fig:sr:parametric_studies:d_1} plots, as a function of $d \ge 2\,\rm \mu m$, the maximum kinetic energy reached by the fastest protons as well as the corresponding THz energy yield. The general trend is that of a monotonic decrease in the ion kinetic energy with the target thickness, from $\Energy_{\rm f} \simeq 15\,\rm MeV$ at $d = 2\,\rm \mu m$ down to $\simeq 5\,\rm MeV$ at $d = 20\,\rm \mu m$. These results are consistent with the measurements reported in Refs.~\onlinecite{McKinnon:prl:2002, Zeil:njp:2010, Poole:njp:2018}. 

\begin{figure*}
    \centering
    \includegraphics[width=0.99\linewidth]{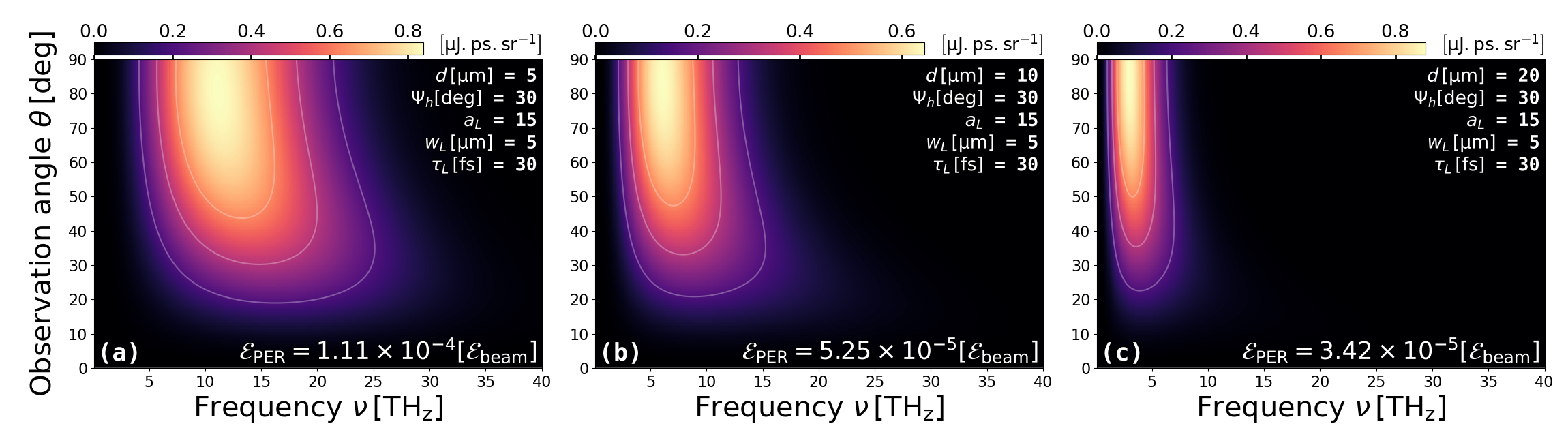}
    \caption{Energy-angle spectra of PER for (a) $5\,\rm \mu m$, (b) $10\,\rm \mu m$ and (c) $20\,\rm \mu m$ thick targets. The laser parameters are those of Fig.~\ref{fig:ctr:electron_beam:radiat_mech_comp}.}
    \label{fig:sr:parametric_studies:d:radiated_energy_spectra_adiab_and_iso}
\end{figure*}

Closely following that trend, the model predicts a continuous drop in the radiated energy from $\Energy_{\rm PER} \simeq 1.7 \times 10^{-4}\,\Energy_{\rm beam} \simeq 95\,\rm \mu J$ at $d=2\,\rm \mu m$ down to $~3.4 \times 10^{-5}\,\Energy_{\rm beam} \simeq 19\,\rm \mu J$ at $d \simeq 20\,\rm \mu m$. Figure~\ref{fig:sr:parametric_studies:aL} further shows that this decrease in $\Energy_{\rm PER}$ at larger $d$ is less pronounced when $a_L$ is reduced: $\Energy_{\rm PER}$ indeed drops by $\sim 5 \times$ at $a_L=15$ and by $\sim 3\times$ at $a_L=5$.

Modifying the thickness of the foil also strongly impacts the frequency-angle spectrum of PER. Figures~ \ref{fig:sr:parametric_studies:d:radiated_energy_spectra_adiab_and_iso}(a)-(c) display the spectra obtained for $d=5~\rm \mu m$, $10~\rm \mu m$ and $20~\rm \mu m$, respectively. The $d=2\,\rm\mu m$ case is depicted in Fig.~\ref{fig:sr:parametric_studies:aL_5_and_aL_15}(b). We notice that, as the target is made thicker, the spectrum, initially distributed in a wide frequency range ($5\,\rm THz \lesssim \nu \lesssim 30 \,\rm THz$), shrinks to a much narrower bandwidth while shifting to smaller frequencies ($1\,\rm THz\lesssim \nu \lesssim 6\,\rm THz$). As expected, though, its angular profile remains unchanged, with a broad maximum around $\sim 70-90\deg$.  

A major prediction of our modeling is that the THz energy radiated from the plasma expansion should be about one to three orders of magnitude lower than that from the fast electrons only. The dominance of the latter is expected to be particularly marked for a fraction of escaping electrons $\eta_{\rm esc} \gtrsim 5\%$ (compare Figs.~\ref{fig:ctr:escaping_electrons:yield} and \ref{fig:sr:parametric_studies:d_1}).


\section{Conclusions}
\label{sec:conclusions}

In summary, we have developed a novel theoretical model of far-field THz emissions in relativistic ultrashort laser-foil interactions, assuming a two-stage scenario. In the first stage, we consider the radiation resulting from the fast electrons alone at the target backside, during their first and brief (lasting only a few fs) excursion into vacuum. In this phase, the vast majority of them are reflected into the target by the sheath field they have themselves created. The THz emission then originates from the combination of coherent transition/synchrotron radiations from the recirculating electrons, as well as from coherent transition radiation from the escaping electrons. In the second stage, all of the confined fast electrons are assumed to have relaxed to a thermal distribution that drives, through the sheath field, the acceleration of the target ions. The dynamics of the nonneutral layers at the edges of the expanding plasma then gives rise to a dipole-type radiation. 

To our knowlege, our unified kinetic model of CTR and CSR, based on the image charge method and integrated over an ensemble of single-particle trajectories, is the first of its kind. Its predictions highlight the critical importance of treating simultaneously these two mechanisms, as they tend to interfere destructively in the THz domain. Their degree of imperfect cancellation, which determines the energy yield and spectrum of the physically meaningful net (CTSR) radiation, depends on the details of the electron trajectories, notably the extent of their excursion in vacuum -- a function of the electron momentum and sheath field strength -- and the amount of escaping (assumed ballistic) electrons.

We have examined the sensitivity of this radiation to the main laser-target parameters, varying these around a reference setup characterized by $a_L=15$, $\tau_L=30\,\rm fs$, $w_L=5\,\rm \mu m$ and $\eta_h=0.2$, corresponding to a (fixed) electron beam energy of $\Energy_{\rm beam}=0.56\,\rm J$. An important finding is that a fraction of escaping electrons as low as $\eta_{\rm esc} \simeq 1\,\%$ -- a value consistent with expectations \cite{Rusby:hplse_2019} -- may suffice to shape the THz spectrum radiated from thin ($d = 2\,\rm \mu m$) foils. At this threshold value, the spectrum is shifted to relatively low frequencies ($\nu \lesssim 5\,\rm Thz$) compared to the case of full electron refluxing, though containing about the same integrated energy ($\simeq 4\times 10^{-3}\,\Energy_{\rm beam} \simeq 2\,\rm mJ$). The latter, however, is predicted to rise tenfold or more should the escaping electron fraction reach $\sim 5\%$. 
The minimum fraction of escaping electrons needed to govern the radiation is an increasing function of the target thickness and degree of laser focusing. For a $20\,\rm \mu m$ thick foil and a $5\,\rm \mu m$ laser spot size ($a_L=15$), this threshold fraction is found to lie between $5\%$ and $10\,\%$, quite a bit above the expected amount ($\sim 1\,\%$) of those electrons \cite{Rusby:hplse_2019}. Consequently, the contribution of the confined electrons, characterized by higher frequencies ($\nu \simeq 5-20\,\rm THz$), may then well dominate the total spectrum when using not-so-thin targets and tightly focused laser pulses. Near the threshold between the two regimes, the THz spectrum will exhibit two distinct bands at low and high frequencies. More generally, the angular distribution of the radiation is broad and tends to peak outside the emission cone of the fast electrons, especially in thin foils or when CTR from escaping electrons prevails. 

Of course, those trends should be considered qualitative as they depend on a number of coupled parameters, which themselves depend on the detailed experimental conditions. In particular, all throughout, we have set to $20\,\%$ the fraction of the laser pulse energy carried by the fast electrons, and to $30\deg$ the angular spread of the latter, even though those quantities are likely affected by the laser intensity and spot size. Moreover, to ensure the tractability of our already computationally heavy model, we have assumed that the fast electrons propagate ballistically across the solid foil (which implies a small enough target thickness) and that the sheath field that confines most of them is both uniform and stationary. Finally, our model being restricted to the first excursion of the fast electrons into vacuum, it discards the subsequent, possibly significant, CTSR-type bursts as the electrons recirculate through the foil and, in so doing, progressively decelerate and scatter.    

These limitations notwithstanding, our modeling indicates that when CTR and CSR insufficiently compensate each other, as occurs for a large enough fraction of escaping electrons or a weak enough sheath field, the radiated energy may approach, or even exceed, the electron beam energy. This result suggests that, under such conditions, one should describe the self-consistent effect of the collective radiation on the electron dynamics. This complex problem can only be quantitatively addressed through three-dimensional, fully kinetic numerical simulations, ideally equipped with a far-field radiation diagnostic \cite{Pardal:cpc:2023}. 

Our model for the THz radiation (PER) emitted at later times as a result of ion acceleration no longer relies on a kinetic treatment of the accelerated particles, because of the daunting complexity of the problem, but rather on a simplified description of the space- and time-varying current density distribution in the expanding plasma. Specifically, we approximate the charge-separation regions to two infinitely thin disks of opposite areal charge, located at the moving ion front and the initial target surface, and describe their dynamics by adapting the plasma expansion model proposed in Ref.~\onlinecite{Ferri:pop:2018}, itself built upon several previous works \cite{Mora:prl:2003, Mora:pre:2005, Brantov:prstab:2015}. Compared to previous attempts at estimating the THz emission from accelerated ions~\cite{Herzer:njp:2018, Woldegeorgis:pre:2019, Liao:pnas:2019, Liao:prx:2020}, our description takes account of the time-varying areal charge of the nonneutral layers and goes beyond the customarily considered isothermal electron limit by considering the transition to the adiabatic regime, as is relevant to micron-thick targets.

Our computations, carried out over parameter ranges similar to those considered for CTSR -- assuming, in particular, the same driving ultrashort laser pulses --, predict that the THz energy radiated during the plasma expansion phase should be at least one order of magnitude below that emitted earlier by the fast electrons. Overall, for the interaction conditions considered, the beam-to-THz conversion efficiency of PER varies in the $\sim 10^{-5}-10^{-4}$ range and is a decreasing function of the target thickness. It also rises with the degree of laser focusing, albeit more and more slowly as the foil is made thicker (for $2\le d \le 20\,\rm \mu m$). As expected, PER produces increasingly low frequencies, and over an increasingly narrow range, when the ion acceleration diminishes, that is, when the target thickness or the laser intensity are reduced. This radiation is mainly emitted at large angles ($>50\deg$), yet this feature alone should not suffice to allow PER to be discerned against the intense background due to CTSR.

Our results evidently call for numerical and experimental validations. Regarding the latter, it should be recalled that most previous experiments have characterized the THz emissions over limited frequency and angular ranges~\cite{Jin:pre:2016, Herzer:njp:2018}, a notable exception being Ref.~\onlinecite{Gopal:pre:2019}, where the full THz distribution was diagnosed at oblique laser incidence. The acquisition of detailed frequency-angle spectra that can be compared with the theoretical spectra would be highly desirable in order to better tune the model parameters and support its predictions.
\\

\appendix

\section{Discrete to continuous description of the energy spectrum}
\label{sec:app:discrete_to_continuous_description}

We detail here the calculation steps between Eq.~\eqref{equ:ctr:electron_beam:radiated_energy_sum_l} and Eq.~\eqref{equ:ctr:electron_beam:radiated_energy_integral} to obtain a continuous description of the energy spectrum radiated by a set of charged particles obeying a given distribution function. We start from the discrete expression of the energy spectrum [see Eq.~\eqref{equ:ctr:electron_beam:radiated_energy_sum_l}] and introduce, for each particle (labeled by the subscript $l$), the amplitude $\A_l$ and the complex phase $\phi_l$ such that
\begin{align}
    \nonumber
    \frac{\partial^2 \SpectralEnergy_{\rm rad}}{\partial \nu \partial \Omega}(\theta, \nu) &= \left\langle \frac{q^2 \nu^2}{2 \eps_0 c} \Biggl| \sum_{l=1}^{N_h} \left[ \A_l^\nameRealParticle (\rhat, \nu) - \A_l^\nameImageParticle (\rhat, \nu) \vphantom{\int} \right] \Biggr|^2 \right\rangle \\
    \label{equ:app:electron_beam:radiated_energy_sum_l_simplified_notation}
    &= \left\langle \frac{q^2 \nu^2}{2 \eps_0 c} \Biggl| \sum_{l=1}^{N_h} \A_l e^{i \phi_l} \Biggr|^2 \right\rangle \,,
\end{align}
where $\langle \cdot \rangle$ indicates an ensemble average. We then introduce a continuous description $\A_{\r,t, \u}(\rhat,\nu)$ and $\phi_{\r,t, \u}(\rhat,\nu)$ of the terms $\A_l$ and $\phi_l$ by imposing that for every particle,
\begin{equation}
    \A_l e^{i \phi_l} = \A_{\r_l,t_l, \u_l}(\rhat,\nu) \exp \left( i \phi_{\r_l,t_l, \u_l}(\rhat,\nu) \right) \,, \nonumber
\end{equation} where $\r_l$, $t_l$ and $\u_l$ represent the position, time and initial momentum of the accelerated particle. The sum of this quantity over $N_h$ particles sampling the $(\r,t,\u)$ space can be expressed as
\begin{widetext}
\begin{align}
    \label{equ:app:electron_beam:radiated_energy:dirac_distributions}
    \Biggl\langle \Biggl| \sum_{l=1}^{N_h} \A_l e^{i \phi_l} \Biggr|^2 \Biggr\rangle  &= \Biggl\langle \Biggl|\, \iiint_{\r, t, \u} \A_{\r,t, \u}(\rhat,\nu) e^{i \phi_{\r,t, \u}(\rhat,\nu)} \times \Biggl( \sum_{l=1}^{N_h} \delta(\r - \r_l) \delta(t - t_l) \delta(\u - \u_l ) \Biggr) \,\dd^2 \r \, \dd t \, \dd^3 \u \Biggr|^2 \Biggr\rangle \,.
\end{align}
\end{widetext}
Next, we introduce the distribution function $F(\r,t,\u)$ that fulfills 
\begin{equation}
    \label{equ:app:electron_beam:dirac_distributions_alone}
    \Biggl\langle \sum_{l=1}^{N_h} \delta(\r - \r_l) \delta(t - t_l) \delta(\u - \u_l ) \Biggr \rangle \to N_h F(\r,t,\u) \,.
\end{equation}
We hereafter assume that the distribution function can be taken in the form of a separable function, $F(\r,t,\u) = f(\r) h(t) g(\u)$.

Finally, we substitute Eq.~\eqref{equ:app:electron_beam:dirac_distributions_alone} into Eq.~\eqref{equ:app:electron_beam:radiated_energy:dirac_distributions} and make use of $\A_{\r,t, \u}(\rhat,\nu) \exp \left( i \phi_{\r,t, \u}(\rhat,\nu) \right) = \A^\nameRealParticle(\rhat, \nu) - \A^\nameImageParticle(\rhat, \nu) \vphantom{\int}$ to obtain the expression of the energy radiated coherently by an ensemble of $N_h \gg 1$ electrons, 
\begin{multline}
    \label{equ:app:electron_beam:radiated_energy:finalized}
    \frac{\partial^2 \SpectralEnergy_{\rm rad}}{\partial \nu \partial \Omega}(\theta, \nu) = \frac{q^2 \nu^2}{2 \eps_0 c} N_h^2  \Biggl| \, \iiint_{\r, t, \u} f(\r) h(t) g(\u)\\
    \times \left[ \A^\nameRealParticle(\rhat, \nu) - \A^\nameImageParticle(\rhat, \nu) \vphantom{\int} \right] \,\dd^2 \r \,\dd t\,\dd^3 \u \Biggr|^2 \,.
\end{multline}
Equation~\eqref{equ:app:electron_beam:radiated_energy:finalized} is identical to Eq.~\eqref{equ:ctr:electron_beam:radiated_energy_integral}.

\section{Spatial Fourier transform of $\partial_t j_\parallel$}
\label{sec:app:Fourier_transform_of_current_density}

The $\rm 1D$ current density defined by Eq.~\eqref{equ:sr:compute_j:compute_j_from_rho:part_2} involves several Dirac delta functions. Its time derivative can be written as a sum of three components:
\begin{align}
\dpa j_\parallel(z,t) / \dpa t = &- \ddot{\sigma}(t) \left[ \mathcal{H}_0 - \mathcal{H}_{z_{\rm f}(t)} \right] \nonumber \\
& - \left[ 2\dot{\sigma}(t) \dot{z}_{\rm f}(t) + \sigma(t') \ddot{z}_{\rm f}(t) \right] \delta_{z_{\rm f}(t)} \nonumber \\
& + \sigma(t) \dot{z}^2_f(t) \delta'_{z_{\rm f}(t)}\,,
\end{align}
where $\dpa \delta(z - z_{\rm f}(t)) / \dpa t = - \dot{z}_{\rm f}(t) \delta'(z-z_{\rm f}(t))$.

The spatial Fourier transform of $\dpa j_\parallel(z,t) / \dpa t$ that appears in the expression of the PER spectrum, Eq.~\eqref{equ:sr:radiated_energy_final}, can then be evaluated analytically upon noting that each of the above components admits a closed-form Fourier transform:
\begin{align}
\label{equ:sr:compute_Fourier_of_j}
& \mathcal{F}\left[ \dpa j_\parallel(z,t) / \dpa t \right](k_z) = \nonumber \\
& \hspace{30pt} - \ddot{\sigma}(t) \mathcal{F}\left[ \mathcal{H}_0 - \mathcal{H}_{z_{\rm f}(t)} \right](k_z) \nonumber \\
& \hspace{30pt} - \left[ 2\dot{\sigma}(t) \dot{z}_{\rm f}(t) + \sigma(t') \ddot{z}_{\rm f}(t) \right] \mathcal{F} \left[ \delta_{z_{\rm f}(t)} \right](k_z) \nonumber \\
& \hspace{30pt} + \sigma(t) \dot{z}^2_{\rm f}(t) \mathcal{F} \left[ \delta'_{z_{\rm f}(t)} \right](k_z)\,,
\end{align}
where
\begin{align}
& \mathcal{F}\left[ \mathcal{H}_0 - \mathcal{H}_{z_{\rm f}(t)} \right](k_z) = \nonumber \\
& \hspace{20pt} z_{\rm f}(t') \, \mathrm{sinc} \left( k_z z_{\rm f}(t') \right) \exp \Bigl( - 2 i \pi k_z z_{\rm f}(t') / 2 \Bigr), \\
& \mathcal{F}\left[{\delta_{z_{\rm f}(t)}(z)} \right](k_z) = \exp \Bigl( - 2 i \pi k_z z_{\rm f}(t') \Bigr), \\
& \mathcal{F}\left[{\delta'_{z_{\rm f}(t)}(z)} \right](k_z) = 2i \pi k_z \exp \Bigl( - 2 i \pi k_z z_{\rm f}(t') \Bigr)\,.
\end{align}
Equation \eqref{equ:sr:compute_Fourier_of_j} is then injected into Eq.~\eqref{equ:sr:radiated_energy_final} where it is evaluated at the wavenumber $k_z = -\nu \cos \theta / c$.

\bibliography{references} 

\end{document}